\documentclass[11pt,a4paper]{article}
\usepackage{jheppub_kim}

\usepackage{pdflscape}
\usepackage{amsmath}
\usepackage{amssymb}
\usepackage{dcolumn}
\usepackage{bm}
\usepackage{color}
\usepackage{epsfig}
\usepackage{amsfonts}
\usepackage{graphicx}
\usepackage{subfigure}
\usepackage{dcolumn}

\newcommand{\be}{\begin{equation}}
\newcommand{\ee}{\end{equation}}
\newcommand{\bea}{\begin{eqnarray}}
\newcommand{\eea}{\end{eqnarray}}

\def\be{\begin{equation}}
\def\ee{\end{equation}}
\def\bea{\begin{eqnarray}}
\def\eea{\end{eqnarray}}

\begin{document}

\title{Thermodynamics of higher dimensional black holes with higher order thermal fluctuations}

\author[a]{B. Pourhassan,}
\author[a]{K. Kokabi,}
\author[a]{S. Rangyan}

 \affiliation[a]{School of Physics, Damghan University, Damghan, 3671641167, Iran}

 \emailAdd{b.pourhassan@du.ac.ir}
  \emailAdd{kokabi@du.ac.ir}
\emailAdd{s.rangyan@std.du.ac.ir}

\abstract{In this paper, we consider higher order corrections of the entropy, which coming from thermal fluctuations, and find their effect on the thermodynamics of higher dimensional charged black holes. Leading order thermal fluctuation is logarithmic term in the entropy  while higher order correction is proportional to the inverse of original entropy. We calculate some thermodynamics quantities and obtain the effect of logarithmic and higher order corrections of entropy on them. Validity of the first law of thermodynamics investigated and Van der Waals equation of state of dual picture studied. We find that five-dimensional black hole behaves as Van der Waals, but higher dimensional case have not such behavior. We find that thermal fluctuations are important in stability of black hole hence affect unstable/stable black hole phase transition.}


\maketitle

\section{Introduction}
Study of the black hole thermodynamics \cite{1410.1486} as well as black hole statistics \cite{9705006, 0807.4520} is one of the important topics of theoretical physics with applications in high energy physics \cite{9408066}, astrophysics and cosmology \cite{1203.2103, 1504.08226}. Black hole thermodynamics is indeed a relation between quantum physics and gravitation \cite{0409024, GRG1, GRG2}. It is indeed Jacobson formalism which relates thermodynamics to the gravity by using the associated Einstein equations with Clausius relation \cite{Jacob}. It is also possible to  obtain the field equations from the equipartition law \cite{Pad}. Interesting fact is that the black hole entropy is proportional to the event horizon area \cite{9303048}, and it is well known as the Bekenstein-Hawking formula \cite{BH1, BH2}. However, this area entropy may be corrected due to the thermal fluctuations or quantum effects. Generally there are two kinds of corrections on the black hole entropy. The first one is power-law corrections which are related to the quantum fields entanglement \cite{Das}. The other corrections arises from loop quantum gravity due to the thermal and quantum fluctuations. These corrections of the entropy are logarithmic at leading order \cite{Log1, Log2, Log3}. These corrections are of the form of $\ln{A}$, where $A$ is the event horizon area. For example, thermodynamics and statistics of G\"{o}del black hole with mentioned logarithmic correction have been studied by the Ref. \cite{Pourdarvish} which is indeed extension of the Ref. \cite{Pourdarvish2}. Also, the generalized second law of thermodynamics in universes with quantum corrected entropy have been studied by the Ref. \cite{1006.3745}. There are other logarithmic corrections as $\ln{S_{0}T^{2}}$ or $\ln{{C_0}T^{2}}$, where $S_{0}$ and $C_{0}$ are original entropy and specific heat respectively, and $T$ is the Hawking temperature \cite{1, 2}. There are several interesting black objects where the effects of thermal fluctuations considered. For example, Hyperscaling violation backgrounds \cite{hyper1, hyper2, hyper3} are of interesting backgrounds where the effect of logarithmic correction investigated recently \cite{hyper4}. The Ref. \cite{hyper4} considered higher order corrections of entropy obtained by the Ref. \cite{higher}. At leading order, effect of logarithmic correction on P-V criticality of AdS black holes in massive gravity \cite{PV1} and dyonic charged AdS black holes \cite{PV2} investigated. Another interesting background is STU model \cite{STU1, STU2, STU3} so effects of the logarithmic correction on the thermodynamics of STU black hole have been studied by the Ref. \cite{STU4}, which may be considered as extension of the Ref. \cite{STU5}. In the case where logarithmic corrections are due to quantum effects, one can investigate quantum gravity effects using holographic principles \cite{QG}. Such thermal fluctuations are important when the size of black hole reduced due to the Hawking radiation, in that case thermodynamics of a infinitesimal singly rotating Kerr-AdS black hole investigated by the Ref. \cite{NPB}.
Black saturns are another interesting thermodynamics objects where the effects of thermal fluctuations investigated \cite{sat1,sat2}. An interesting kinds of regular black holes which have not contain a singularity at the center are Hayward black hole \cite{hay1} so the effects of thermal fluctuations at leading order on the thermodynamics of modified Hayward black hole have been studied by the Ref. \cite{hay2}. Moreover, thermal fluctuations in a charged AdS black hole have been considered by the Ref. \cite{1503.07418} and have shown that affect stability of black hole. There are also some recent works to show effects of thermal fluctuations on the black holes thermodynamics \cite{1701, 1705}. These show importance of consideration of thermal fluctuations. Hence, in this paper we would like to use higher order corrected entropy \cite{higher} in the thermodynamics of higher dimensional black holes. Ref. \cite{higher} obtained general form of higher order correction of the entropy and applied it to the BTZ black hole \cite{9204099, 0903.0292, 1611}, and the Anti
de-Sitter Schwarzschild black hole. In this work, we are interested to the higher dimensional black hole which are responsible for M-theory, string theory and AdS/CFT \cite{Mal}. As we know, M-theory admit 11-dimensional space-time \cite{9711053}. Moreover, string theory require 10-dimensional space-time, hence 10-dimensional black holes may be of our interest. Considering black holes in string theory show that the area law black hole entropy may be arises from counting microscopic configurations \cite{9607235}. Finally, we should note that AdS/CFT as well as AdS/CMT \cite{CMT} and AdS/QCD require 5-dimensional black hole. Hence, higher dimensional black holes are important and interesting subjects in theoretical physics \cite{0801.3471}. Myers-Perry solutions are of most important five-dimensional black hole \cite{0308056} which could be considered for thermodynamics and statistical analysis \cite{Pourdarvish3}. Also, Reissner-Nordstr\"{o}m black hole, which is a charged black hole, in higher dimensions is interesting kind of black hole, because it has several similarities with Schwarzschild black holes \cite{1709}. In that case, stability of higher dimensional Reissner-Nordstr\"{o}m-anti-de Sitter black holes investigated by the Ref. \cite{0809.2048} and found that there are some unstable thermodynamics regions. Similar instability also discussed by the Refs. \cite{1309.7667} and  \cite{1511.06059}.  The area quantum of the higher dimensional Reissner-Nordstr\"{o}m black hole in the small black hole charge have been studied by the Ref. \cite{1003.4248}. In presence of instability, it may be possible to have phase transition, which have been studied by the Ref. \cite{1502.01428} for the 5-dimensional Reissner-Nordstr\"{o}m black hole. These give us strong motivation to study thermodynamics of higher dimensional Reissner-Nordstr\"{o}m black holes in presence of correction terms of the entropy.\\
This paper is organized as follows. In the next section we review origin of the corrected entropy which will be considered in this paper. In section 3 we review some important aspects of Reissner-Nordstr\"{o}m black hole in $d$-dimensional space-time. Then in section 4 we study corrected thermodynamics and obtain some thermodynamics quantities like Helmholtz and Gibbs free energies and enthalpy. In section 5 we investigate validity of the first law of thermodynamics. Then, in the section 6 we discuss about possible dual picture of the black hole, and in the section 7 we study about critical points and stability of the black hole under effects of higher order corrections of the entropy. Finally, in section 8 we give conclusion and some outlook for the future works.
\section{Higher order thermal fluctuations}
Ref. \cite{higher} found general expression for the higher order corrected entropy. In that case the quantum density of the system is
then given by,
\begin{equation}\label{1}
\rho(E)=\sum_{n}{\Omega(E_n)\delta(E-E_n)},
\end{equation}
where $\Omega(E_{n})$ denotes number of microstates at energy $E_{n}$, while $E$ is the average energy of the given system. Also, $\delta(E-E_n)$ is the delta function. It yields to the following partition function of $N$ particle system in the canonical ensemble,
\begin{equation}\label{2}
Z_{c}(\beta)=\sum_{n}{\Omega(E_{n})\exp (-\beta E_{n})}=\int_{0} ^{\infty}{\rho(E)\exp (-\beta E)dE},
\end{equation}
where index $c$ in the $Z_{c}$ denotes partition function in the canonical ensemble, and $\rho(E)$ given by the equation (\ref{1}). Also, $\beta=\frac{1}{T}$ in unit of Boltzmann constant. Then, by using the Laplace transform of the density of states we have,
\begin{equation}\label{3}
\rho(E)=\frac{1}{2\pi i}\int^{i\infty}_{-i\infty}{e^{S(\beta)}d\beta},
\end{equation}
where
\begin{equation}\label{4}
S(\beta)=\ln (Z_{c}(\beta))+\beta E.
\end{equation}
Now, one can use Taylor expansion around equilibrium temperature $\beta_{0}$ to find,
\begin{eqnarray}\label{5}
S(\beta)&=&S_{0}(E)+\left(\frac{\partial}{\partial\beta}\ln Z_{c}\right)_{\beta_{0}}(\beta -\beta_{0})\nonumber\\
&+&\frac{1}{2!}\left(\frac{\partial^{2}}{\partial\beta^{2}}\ln Z_{c}\right)_{\beta_{0}}(\beta -\beta_{0})^{2}+\frac{1}{3!}\left(\frac{\partial^{3}}{\partial\beta^{3}}\ln Z_{c}\right)_{\beta_{0}}(\beta - \beta_{0})^{3}+\cdots.
\end{eqnarray}
It is obvious that the first derivative (last term of the first line) at equilibrium is zero. Hence, the equation (\ref{3}) reduced to the following expression,
\begin{equation}\label{6}
\rho(E)=\frac{1}{2\pi i}\int^{i\infty}_{-i\infty}{e^{S_{0}(E)+\frac{1}{2}\alpha^{2}_{2}(\beta-\beta_{0})^{2}
+\frac{1}{3!}\alpha_{3}(\beta - \beta_{0})^{3}+\frac{1}{4!}\alpha_{4}(\beta - \beta_{0})^{4}+\cdots}d\beta},
\end{equation}
where coefficients defined as follows,
\begin{equation}\label{7}
\alpha^{2}_{2}=\left(\frac{\partial^{2}}{\partial \beta^{2}}\ln{Z_{c}}\right)_{\beta_{0}},
\end{equation}
and
\begin{equation}\label{8}
\alpha_{n}=\left(\frac{\partial^{n}}{\partial\beta^{n}}\ln{Z_{c}}\right)_{\beta_{0}}.
\end{equation}
The second derivative separated because it assumed as positive quantity. Then, by using the $(\beta-\beta_{0})\rightarrow iy$ change of variable and exponential expansion one can obtain,
\begin{equation}\label{9}
\rho(E)=\frac{1}{2\pi}e^{S_{0}}\int^{\infty}_{-\infty} e^{-\frac{1}{2}\alpha^{2}_{2}y^{2}}\left(\sum^{\infty}_{m=0}\frac{1}{m!}(\sum^{\infty}_{n=3}\frac{\alpha_{n}(i)^{n} y^{n}}{n!})^{m}\right),
\end{equation}
Finally, we obtain the integral expression for $\rho(E)$ as follow,
\begin{equation}\label{10}
\rho(E)=\frac{1}{2\pi}e^{S_{0}}\sqrt{\frac{2\pi}{\alpha^{2}_{2}}}\left[1+\sum^{\infty}_{n=2}\frac{\alpha_{2n}(-1)^{n}}{(2n)!!\alpha^{2n}_{2}}
+\frac{1}{2!}\sum^{\infty}_{n=3}\sum^{\infty}_{m=3}\frac{\alpha_{n}\alpha_{m}(-1)^{k}(2k-1)!!}{n!m!\alpha^{2k}_{2}}+\cdots\right].
\end{equation}
Therefore, one can obtain,
\begin{eqnarray}\label{11}
S(E)&=&S_{0}(E)-\ln(n)-\frac{1}{2}\ln(\alpha^{2}_{2})\nonumber\\
&+&\ln\left(1+\sum^{\infty}_{n=2}\frac{\alpha_{2n}(-1)^{n}}{(2n)!!\alpha^{2n}_{2}}
+\frac{1}{2!}\sum^{\infty}_{n=3}\sum^{\infty}_{m=3}\frac{\alpha_{n}\alpha_{m}(-1)^{k}(2k-1)!!}{n!m!\alpha^{2k}_{2}}+\cdots\right),
\end{eqnarray}
The last line approximated by $\ln(1+x)\sim x$ for the small $x$ to find,
\begin{eqnarray}\label{12}
S(E)&=&S_{0}(E)-\ln(n)-\frac{1}{2}\ln(\alpha^{2}_{2})\nonumber\\
&+&\sum^{\infty}_{n=2}\frac{\alpha_{2n}(-1)^{n}}{(2n)!!\alpha^{2n}_{2}}
+\frac{1}{2!}\sum^{\infty}_{n=3}\sum^{\infty}_{m=3}\frac{\alpha_{n}\alpha_{m}(-1)^{k}(2k-1)!!}{n!m!\alpha^{2k}_{2}}+\cdots.
\end{eqnarray}
In order to find useful expression which will be applicable easy for any kinds of black objects we use the following relation,
\begin{equation}\label{13}
S(\beta)=a\beta^m +\frac{b}{\beta^n},
\end{equation}
where $a$ and $b$ are arbitrary constant. The relation (\ref{13}) is consistent with the assumption that the black hole is Euclidean. Then, assume equilibrium point as,
\begin{equation}\label{14}
\beta_{0}=(\frac{nb}{ma})^\frac{1}{m+n}.
\end{equation}
After some calculations and expanding around equilibrium point one can obtain the following general expression,
\begin{equation}\label{15}
S=S_{0}-\frac{\alpha}{2}\ln(S_{0}T^{2})+\frac{\gamma}{S_{0}},
\end{equation}
where $\alpha$ and $\gamma$ are coefficients of corrections. Usual value of $\alpha$ is unit while $\gamma$ may has negative value, however we can choose general value to find effects of correction terms. It is clear that leading order correction is logarithmic, while the second order correction is proportional to the inverse of original entropy. The fact is that the logarithmic term is due to thermal fluctuations, and the second order correction arises when we extend the entropy function around the equilibrium point. In this paper, we would like use corrected entropy (\ref{15}) to study the thermodynamics of higher dimensional Reissner-Nordstr\"{o}m black hole.

\section{Higher dimensional Rissner-Nordstr\"{o}m black hole}
One of the important kinds of black hole is charged non-rotating black hole known as Rissner-Nordstr\"{o}m black hole \cite{0203049}. In that case the $d$-dimensional Rissner-Nordstr\"{o}m black hole is represented by the following metric \cite{0703231, 0804.0295, 0903.1983},
\begin{equation}\label{metric}
ds^2=-f(r)dt^{2}+\frac{dr^{2}}{f(r)}+r^{2}d\Omega_{d-2}^{2},
\end{equation}
where natural units is used ($G=c=\hbar=1$) and,
\begin{equation}\label{16}
f(r)=1-\frac{16\pi M}{(d-2)\Omega_{d-2}r^{d-3}}+\frac{8 \pi Q^{2}}{(d-2)(d-3)\Omega_{d-2}^{2}r^{2(d-3)}},
\end{equation}
also, $M$ and $Q$ are black hole mass and electric charge respectively. Moreover, where $\Omega_{d-2}$ is volume of unit $(d-2)$-sphere given by,
\begin{equation}\label{17}
\Omega_{d-2}=\frac{\pi^{\frac{d}{2}}}{(\frac{d}{2})!}.
\end{equation}
Horizon structure obtained by $f(r)=0$ which yields to the following real positive roots as inner and outer horizons \cite{0903.1983},
\begin{equation}\label{18}
r_{\pm}=\left(\frac{8\pi}{(d-2)\Omega_{d-2}}\left[M\pm\sqrt{M^{2}-\frac{(d-2)Q^{2}}{8\pi(d-3)}}\right]\right)^{\frac{1}{d-3}}.
\end{equation}
We can see that the black hole event horizon is depend on the black hole mass and electric charge as well as space-time dimensions. It is clear that,
\begin{equation}\label{19}
M^{2}-\frac{(d-2)Q^{2}}{8\pi(d-3)}>0,
\end{equation}
to have separated inner and outer horizon, while
\begin{equation}\label{20}
M^{2}=\frac{(d-2)Q^{2}}{8\pi(d-3)},
\end{equation}
is extremal case where $r_{+}=r_{-}$. Otherwise we have only naked singularity. Physical properties of black holes like Hawking temperature and Bekenstein-Hawking entropy given at outer horizon $r_{+}$.\\
Hawking temperature is given by,
\begin{equation}\label{21}
T=\left[\frac{(d-3)}{4\pi}\left(\frac{\Omega_{d-2}}{4}\right)^{\frac{1}{d-2}}\right]S_{0}^{\frac{-1}{d-2}}\left(1-2\left[\frac{(d-2)\Omega_{d-2}Q}{(d-3) 16 \pi M }\right]^{2}\right),
\end{equation}
where $S_{0}$ is the Bekenstein-Hawking entropy of black hole which is given by,
\begin{equation}\label{22}
S_{0}=\frac{\Omega_{d-2}r_{+}^{d-2}}{4},
\end{equation}
hence, black hole temperature is proportional to $\frac{1}{r_{+}}$.
In the next section we use corrected entropy (\ref{15}) to extract some thermodynamics quantities.

\section{Corrected thermodynamics}
It is expected that thermal fluctuations modified thermodynamics of black holes. It may affect also stability of black holes. Hence, first in this section we calculate some important thermodynamics quantities. We are interested to four (ordinary dimensions of our universe), five (from AdS/CFT point of view), ten (inspired from superstring theory) and eleven (responsible for M-theory) space-time dimensions. In all cases we assume that electric charge of the black hole is constant and fins all results in units of $Q$ and consider $M$ as only free parameter of the model. Also we will take $\alpha=1$ as expected from mathematic and analyze some values of $\gamma$.

\subsection{Helmholtz free energy}
\begin{figure}[h!]
 \begin{center}$
 \begin{array}{cccc}
\includegraphics[width=70 mm]{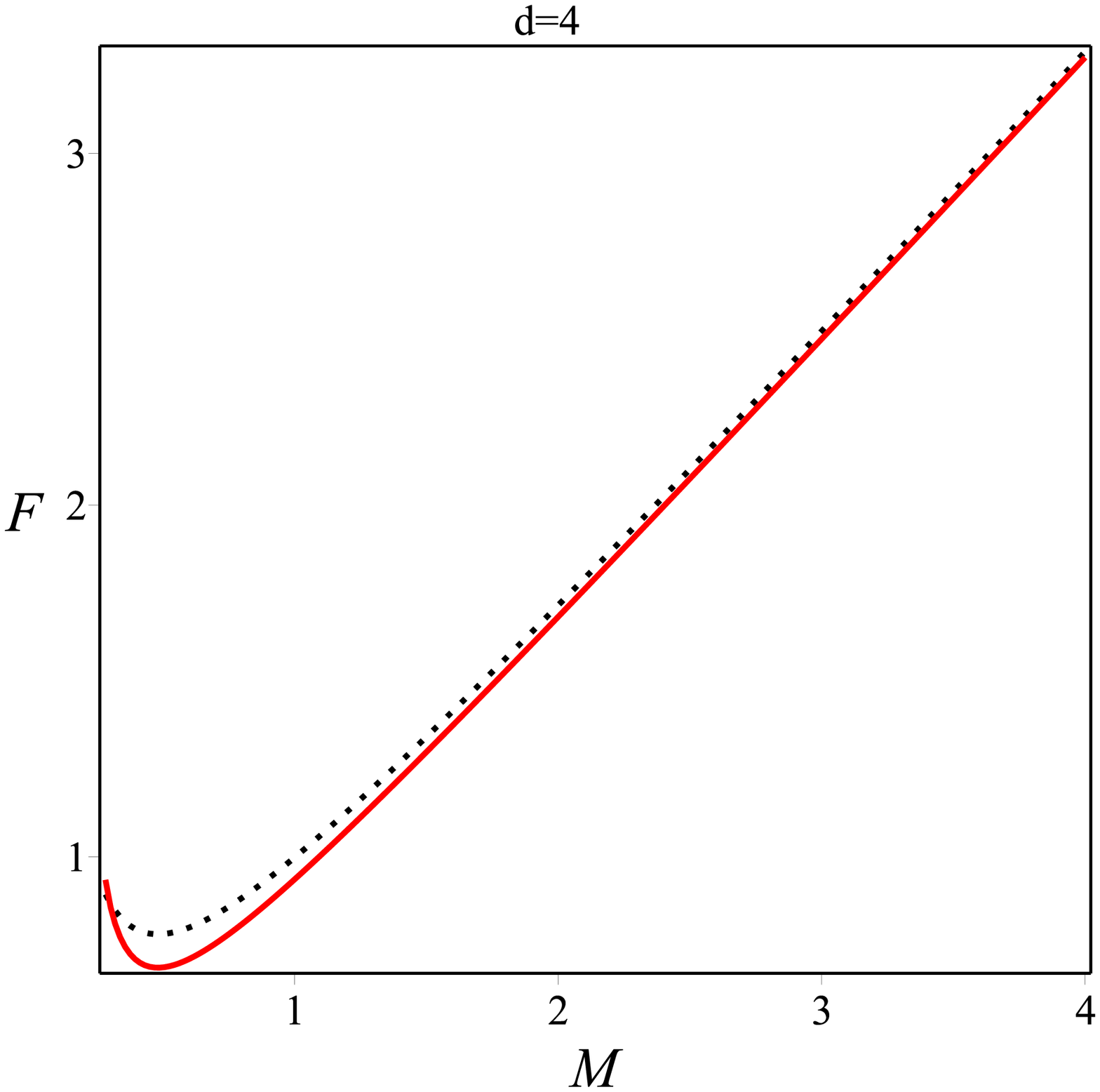}\includegraphics[width=70 mm]{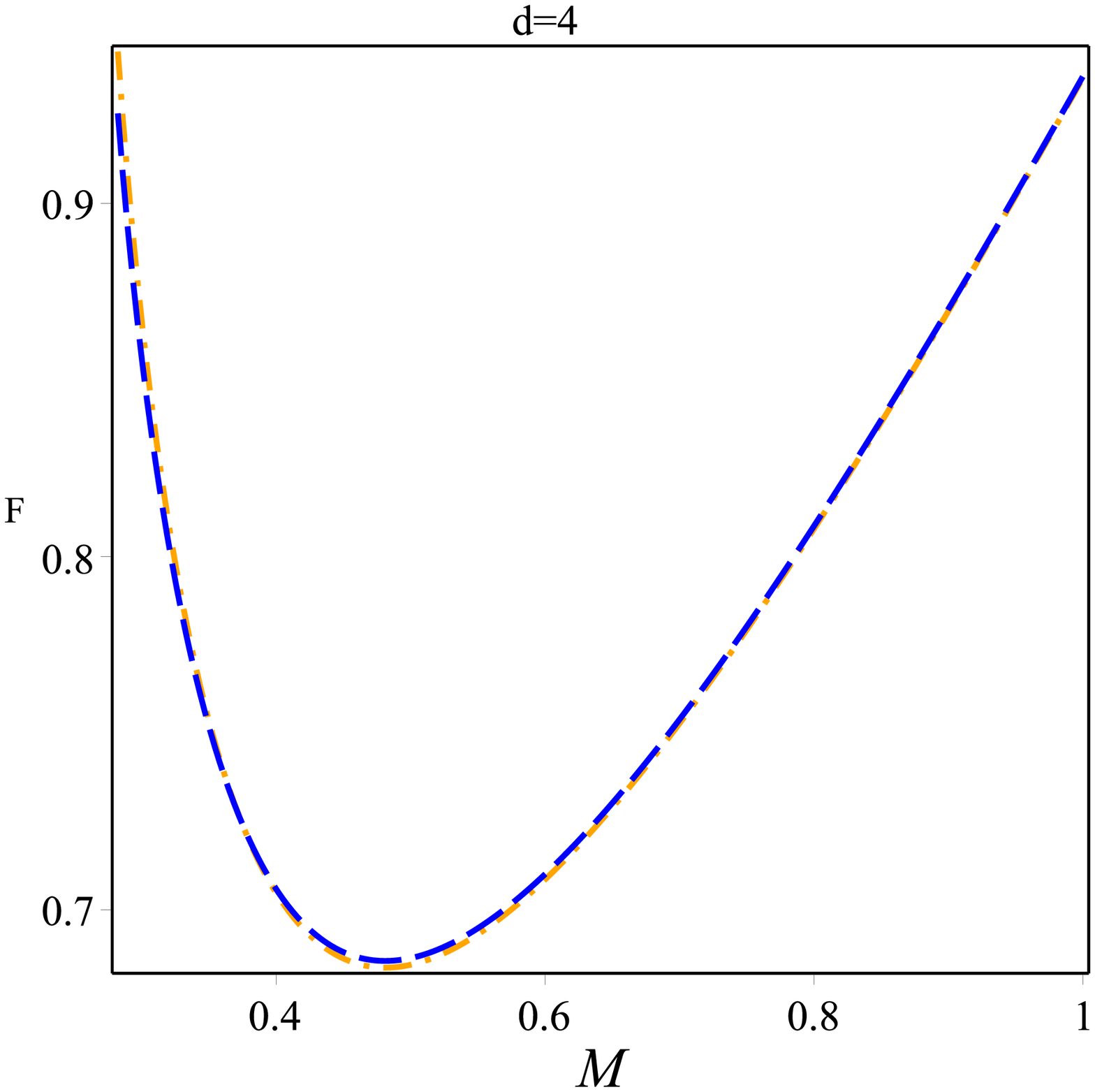}\\
 \end{array}$
 \end{center}
\caption{Helmholtz free energy in terms of $M$ for $d=4$.
$\alpha=0$ and $\gamma=0$ (black dot); $\alpha=1$ and $\gamma=0$ (red solid); $\alpha=1$ and $\gamma=-1$ (blue dash); $\alpha=1$ and $\gamma=1$ (orange dash dot).}
 \label{fig1}
\end{figure}

Helmholtz free energy obtained using the following relation,
\begin{equation}\label{23}
F=-\int {SdT},
\end{equation}
where $T$ given by the equation (\ref{21}) and $S$ given by the equation (\ref{15}). In the case of $d=4$ and $Q=1$ we know from the condition (\ref{19}) that $M>0.2821$. By using the equation (\ref{23}) one can obtain,
\begin{eqnarray}\label{23}
F(d=4)&\approx&\frac{3}{16M}+\frac{8M}{\pi^{2}}\nonumber\\
&+&\frac{\alpha}{M^{3}}\left[\frac{15}{1024}\ln{2}+\frac{\ln{\pi}}{256}+\frac{1}{768}-\frac{1}{1024}\ln{(\frac{\pi^{2}}{M^{2}}-128)^{2}}\right]\nonumber\\
&-&\frac{\alpha}{M}\left[\frac{15}{8\pi^{2}}\ln{2}-\frac{\ln{\pi}}{2\pi^{2}}+\frac{1}{8\pi^{2}}\ln{(\frac{\pi^{2}}{M^{2}}-128)^{2}}\right]\nonumber\\
&+&\frac{\gamma}{M^{3}}\left(\frac{3}{81920M^{2}}-\frac{1}{384\pi^{2}}\right).
\end{eqnarray}
In the plots of the Fig. \ref{fig1} we can see behavior of the Helmholtz free energy (at $d=4$) in terms of $M$ to find effects of correction terms of entropy.
We can find that effect of logarithmic correction as well as higher order correction is decreasing value of the Helmholtz free energy.
Most of corrections effects are at small $M$ which tells that effect of thermal fluctuations are important when size of black hole reduced due to the
Hawking radiation. Also, we can see that Helmholtz free energy has a minimum which affected by logarithmic correction.
We also find that effect of higher order corrections are infinitesimal.

For further dimensions, we examine Helmholtz free energy for five, ten and eleven dimensions.\\
In the case of $d=5$ and $Q=1$ we know from the condition (\ref{19}) that $M>0.24431$, and find following expression for the Hallmoltz free energy at five dimensions,
\begin{eqnarray}\label{24}
F(d=5)&\approx&\frac{15\times10^{-5}}{M^{4}}\left[1.8\times10^{5}M^{5}+0.45\times10^{5}M^{3}\right]\nonumber\\
&+&\frac{15\times10^{-5}\alpha}{M^{\frac{5}{2}}}\left[(0.25\times10^{5}M^{2}-0.012\times10^{5})
\ln{\left(\frac{(5655M^{2}-279)^{2}}{M^{\frac{7}{2}}}\right)}\right]\nonumber\\
&-&\frac{15\times10^{-5}\alpha}{M^{\frac{5}{2}}}\left[4.7\times10^{5}M^{2}-0.26\times10^{5}\right]\nonumber\\
&-&\frac{15\times10^{-5}\gamma}{M^{4}}\left[1618M^{2}-200\right].
\end{eqnarray}

\begin{figure}[h!]
 \begin{center}$
 \begin{array}{cccc}
\includegraphics[width=70 mm]{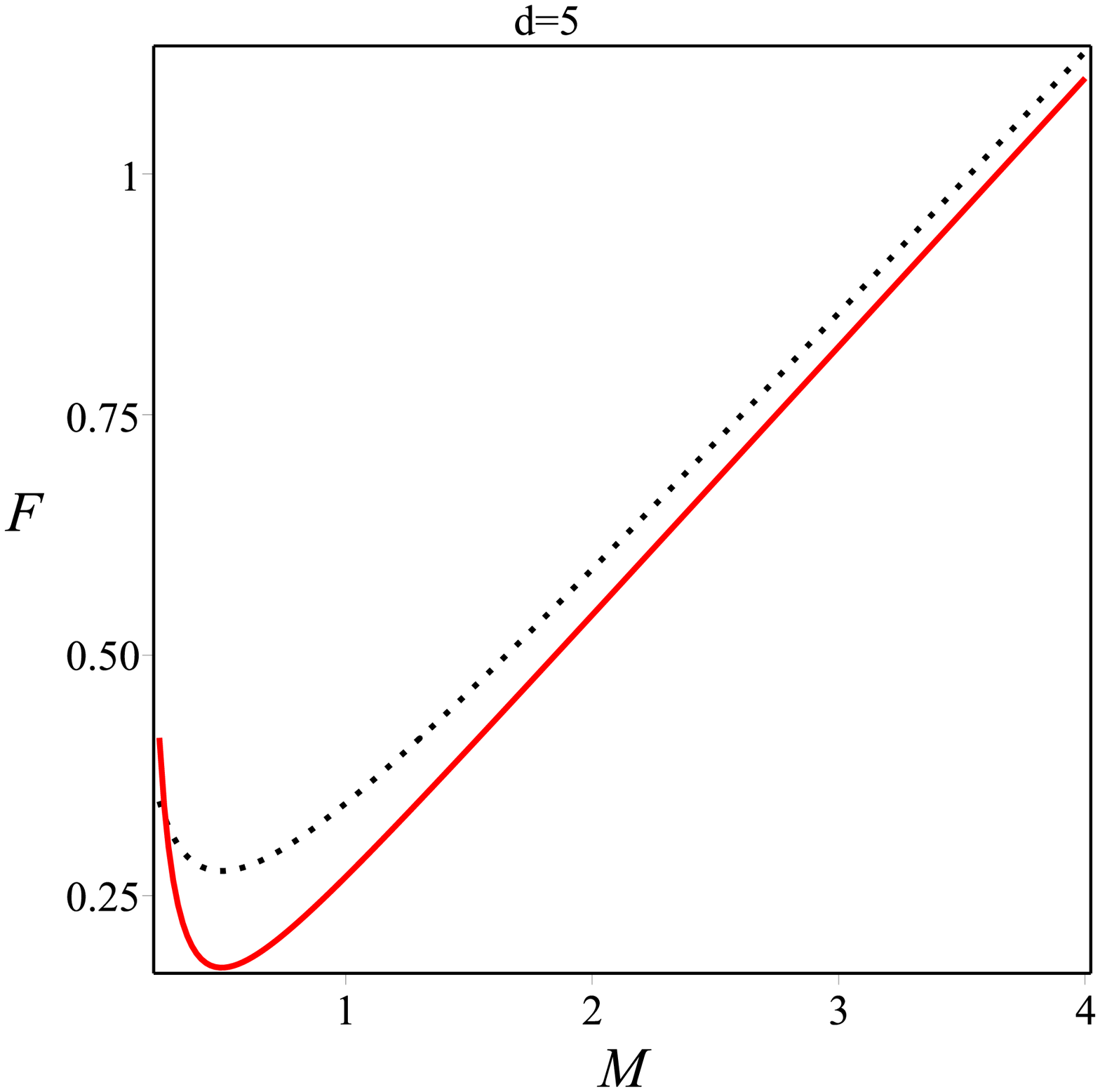}&\includegraphics[width=70 mm]{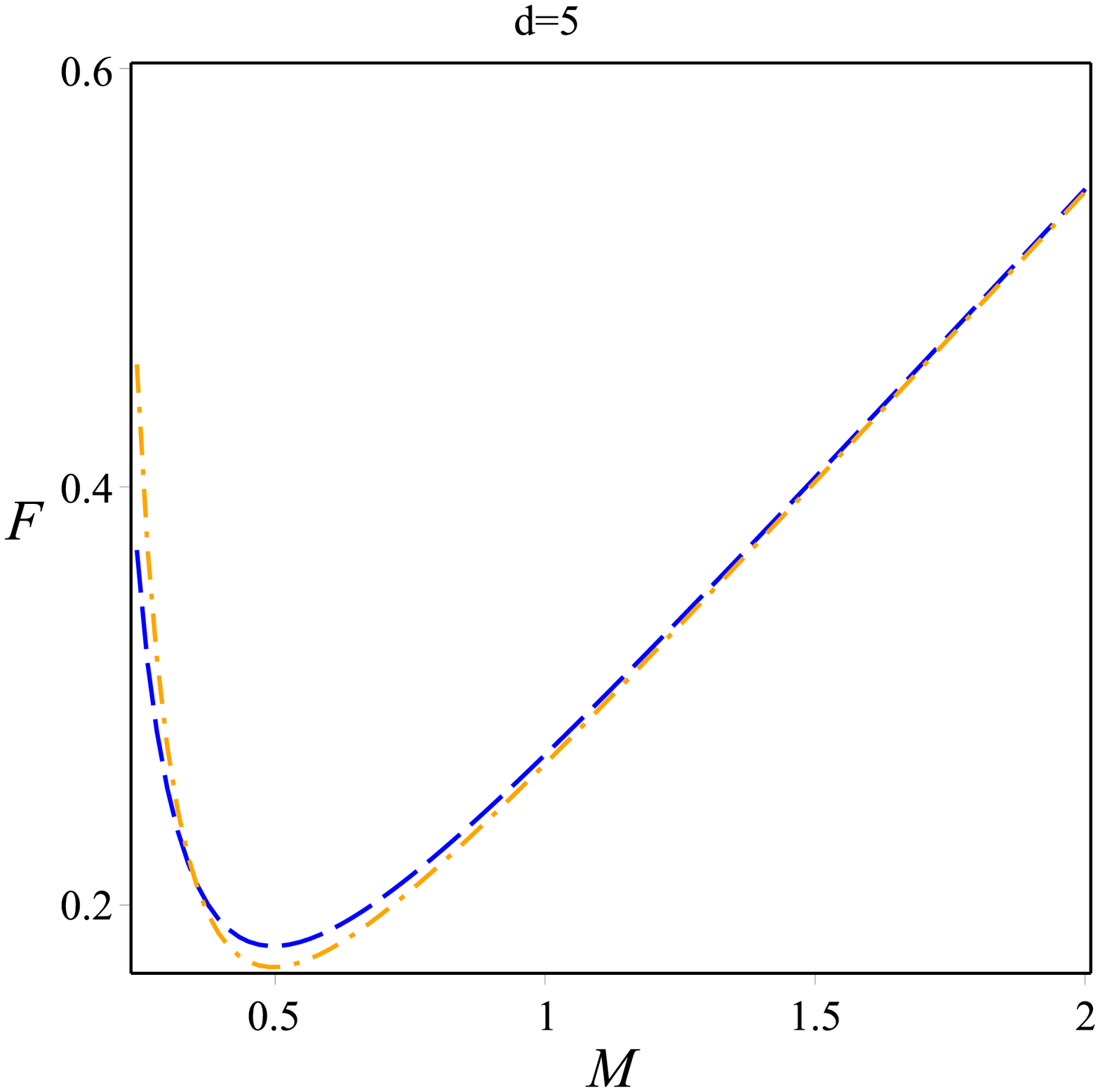}\\
 \end{array}$
 \end{center}
\caption{Helmholtz free energy in terms of $M$ for $d=5$.
$\alpha=0$ and $\gamma=0$ (black dot); $\alpha=1$ and $\gamma=0$ (red solid); $\alpha=1$ and $\gamma=-1$ (blue dash); $\alpha=1$ and $\gamma=1$ (orange dash dot).}
 \label{fig2}
\end{figure}

In the plots of the Fig. \ref{fig2} we can see typical behavior of the Helmholtz free energy (at $d=5$) with respect to $M$ and find similar correction terms effects as the case of four dimensional. It means that the effect of logarithmic correction as well as higher order correction is decreasing value of the Helmholtz free energy at $d=5$. Also, for the larger $M$ one can find that effects of thermal fluctuations are negligible.

\begin{figure}[h!]
 \begin{center}$
 \begin{array}{cccc}
\includegraphics[width=80 mm]{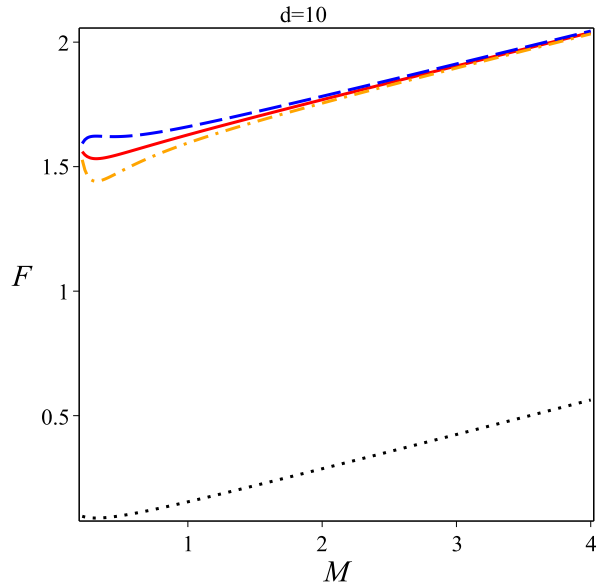}
 \end{array}$
 \end{center}
\caption{Helmholtz free energy in terms of $M$ for $d=10$.
$\alpha=0$ and $\gamma=0$ (black dot); $\alpha=1$ and $\gamma=0$ (red solid); $\alpha=1$ and $\gamma=-1$ (blue dash); $\alpha=1$ and $\gamma=1$ (orange dash dot).}
 \label{fig3}
\end{figure}

In the case of $d=10$ and $Q=1$ we know from the condition (\ref{19}) that $M>0.21324$, and find different behavior for the Helmholtz free energy at ten dimensions which is illustrated by the Fig. \ref{fig3}. Opposite of the $d=4$ and $d=5$ cases we can see that correction terms increase value of the Helmholtz free energy. Also, we can see non-trivial effect of thermal fluctuations for the large and massive black hole. In the case of higher order correction with negative coefficient (as expected) we can see that there is an unstable extremum for the Helmholtz free energy at small $M$ which reflects an instability for the infinitesimal black hole. Later, we will discuss about such instabilities by using specific heat.\\

\begin{figure}[h!]
 \begin{center}$
 \begin{array}{cccc}
\includegraphics[width=80 mm]{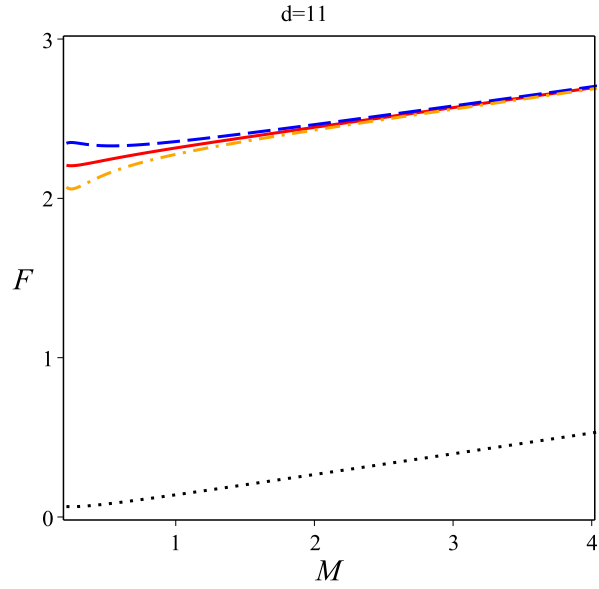}
 \end{array}$
 \end{center}
\caption{Helmholtz free energy in terms of $M$ for $d=11$.
$\alpha=0$ and $\gamma=0$ (black dot); $\alpha=1$ and $\gamma=0$ (red solid); $\alpha=1$ and $\gamma=-1$ (blue dash); $\alpha=1$ and $\gamma=1$ (orange dash dot).}
 \label{fig4}
\end{figure}

In the case of $d=11$ and $Q=1$ we know from the condition (\ref{19}) that $M>0.21157$. In that case, as illustrated by the Fig. \ref{fig4}, we can see similar behavior with $d=10$ case as expected. We can see that higher order corrections are negligible for larger $M$. We can see that Helmholtz free energy has a minima at leading order of correction. Hence, we show that behavior of the Helmholz free energy at $d=4 ,5$ are different with $d=10, 11$ cases.\\

\begin{figure}[h!]
 \begin{center}$
 \begin{array}{cccc}
\includegraphics[width=80 mm]{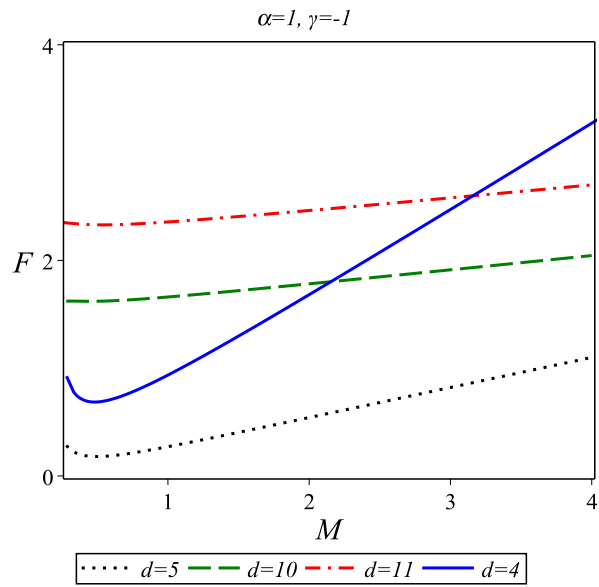}
 \end{array}$
 \end{center}
\caption{Helmholtz free energy in terms of $M$ for different dimensions.}
 \label{fig5}
\end{figure}

In order to compare corrected Helmholtz free energy in several dimensions we give plot of the Fig. \ref{fig5}.

\subsection{Internal energy}
\begin{figure}[h!]
 \begin{center}$
 \begin{array}{cccc}
\includegraphics[width=70 mm]{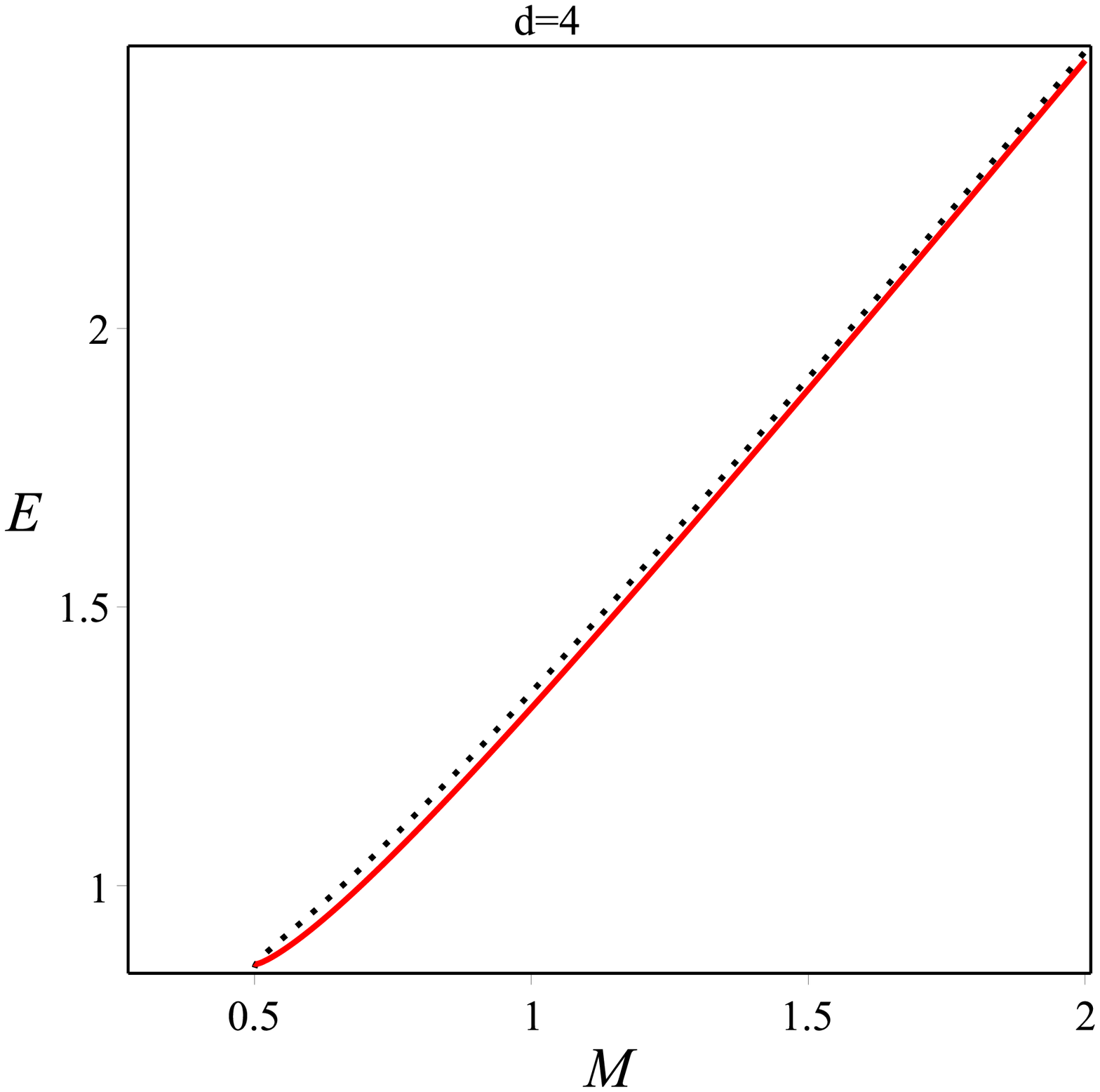}\includegraphics[width=70 mm]{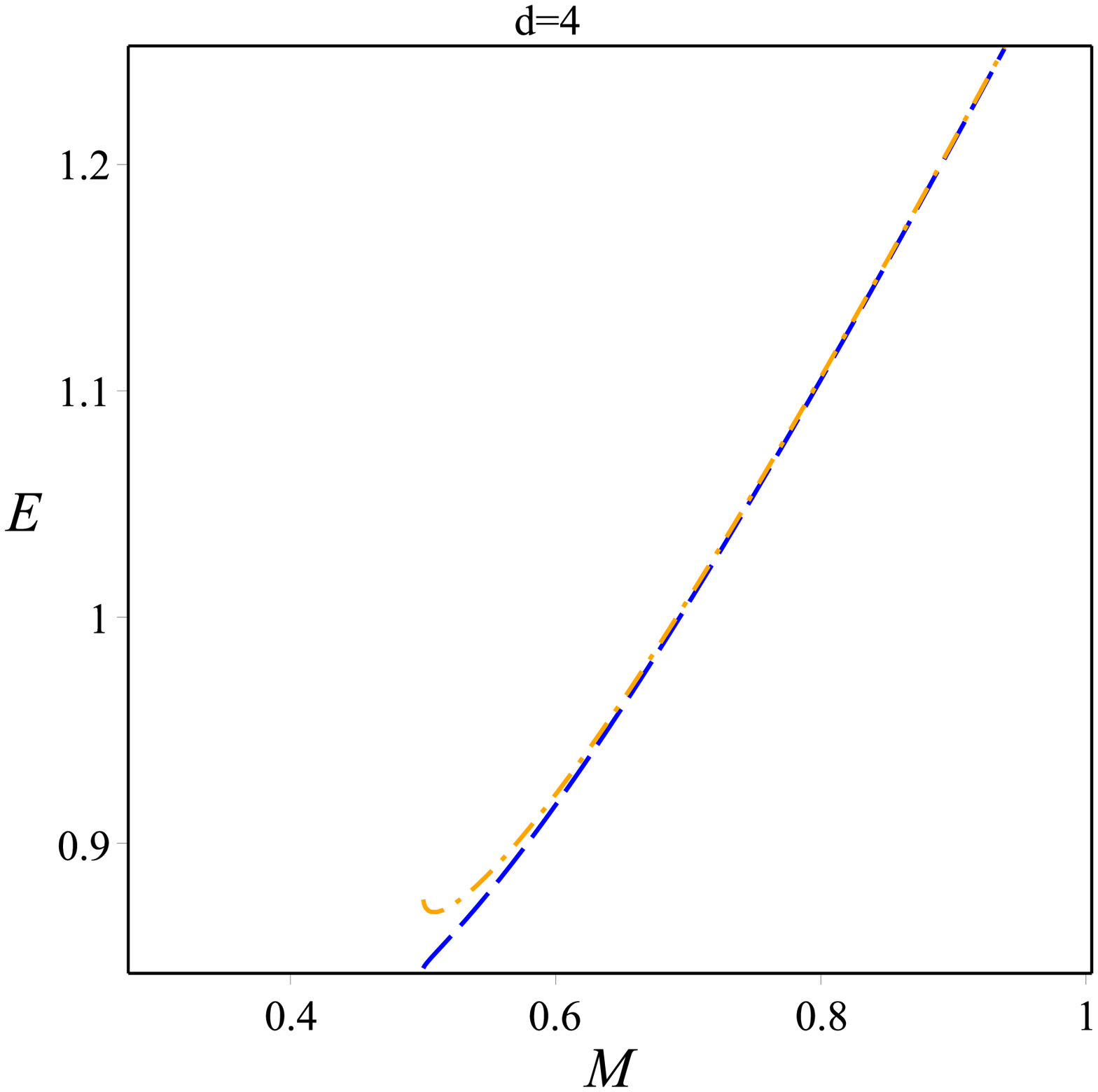}
 \end{array}$
 \end{center}
\caption{Internal energy in terms of $M$ for $d=4$.
$\alpha=0$ and $\gamma=0$ (black dot); $\alpha=1$ and $\gamma=0$ (red solid); $\alpha=1$ and $\gamma=-1$ (blue dash); $\alpha=1$ and $\gamma=1$ (orange dash dot).}
 \label{fig6}
\end{figure}

Internal energy is one of the important thermodynamics quantity. In order to obtain the internal energy of the given system we use the following thermodynamics relationship,
\begin{equation}\label{25}
E=F+ST.
\end{equation}
As before, we try to obtain behavior of internal energy for $d=4$, $d=5$, $d=10$ and $d=11$ space-time dimensions.
In the case of $d=4$ we yield to the following expression,
\begin{eqnarray}\label{26}
E(d=4)&\approx&\frac{384M^{2}+5\pi^{2}}{32M\pi^{2}}\nonumber\\
&+&\frac{\alpha}{1024\pi^{2}M^{3}}\left(128M^{2}-\pi^{2}\right)\left(\ln{(28M^{2}-\pi^{2})}-2\ln{M}\right)\nonumber\\
&-&\frac{\alpha}{6144\pi^{2}M^{3}}\left[(5760\ln{2}+1526\ln{\pi})M^{2}-(8+12\ln{\pi}+45\ln{2})\pi^{2}\right]\nonumber\\
&+&\frac{\gamma}{M^{5}}\frac{3640M^{2}+3\pi^{2}}{491520\pi^{2}}.
\end{eqnarray}
In the plots of the Fig. \ref{fig6} we can see typical behavior of the internal energy (at $d=4$) in terms of $M$ and find effects of correction terms of entropy.
We can find that effect of both leading order (logarithmic) correction and higher order correction are decreasing value of the internal energy. From the right plot of the Fig. \ref{fig6} we can see that difference of positive and negative coefficients of higher order correction is at finite range of $M$. We find that effect of correction terms on the internal energy at four dimension is infinitesimal.\\

\begin{figure}[h!]
 \begin{center}$
 \begin{array}{cccc}
\includegraphics[width=80 mm]{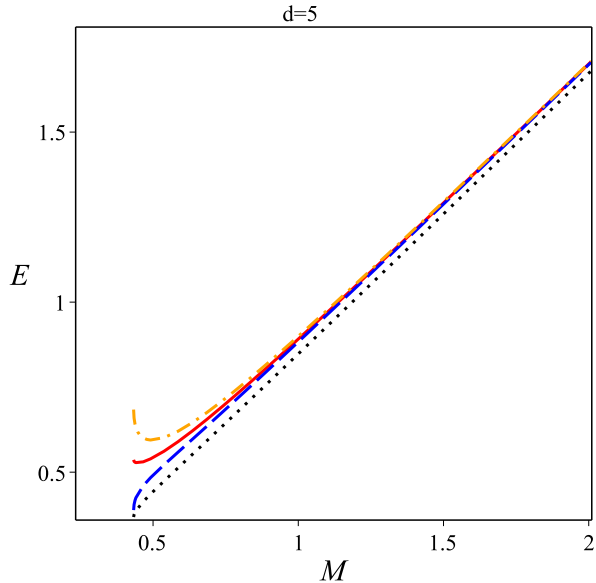}
 \end{array}$
 \end{center}
\caption{Internal energy in terms of $M$ for $d=5$.
$\alpha=0$ and $\gamma=0$ (black dot); $\alpha=1$ and $\gamma=0$ (red solid); $\alpha=1$ and $\gamma=-1$ (blue dash); $\alpha=1$ and $\gamma=1$ (orange dash dot).}
 \label{fig7}
\end{figure}

In the case of $d=5$ we can see different behavior. As illustrated by the Fig. \ref{fig7}, the effect of correction terms are increasing of internal energy. In the case of higher order correction with positive coefficient (dash dot orange curve of the Fig. \ref{fig7}) we can see a minimum for the internal energy. Otherwise it is approximately increasing linear function of $M$. In all cases there is a minimum value for the black hole mass to have internal energy.\\

\begin{figure}[h!]
 \begin{center}$
 \begin{array}{cccc}
\includegraphics[width=80 mm]{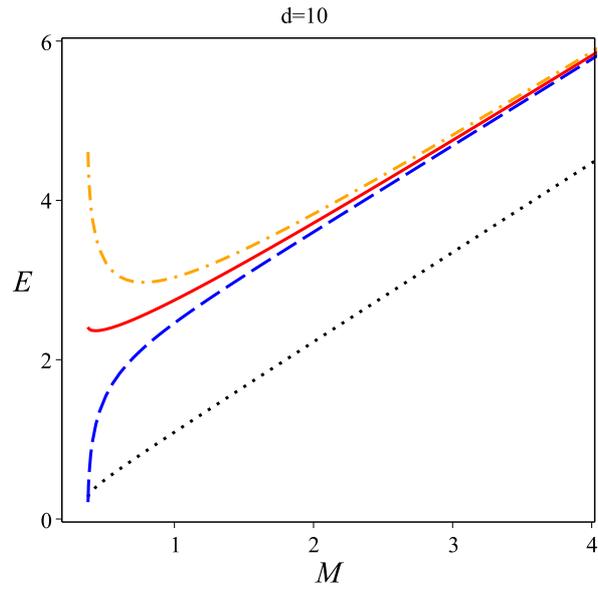}
 \end{array}$
 \end{center}
\caption{Internal energy in terms of $M$ for $d=10$.
$\alpha=0$ and $\gamma=0$ (black dot); $\alpha=1$ and $\gamma=0$ (red solid); $\alpha=1$ and $\gamma=-1$ (blue dash); $\alpha=1$ and $\gamma=1$ (orange dash dot).}
 \label{fig8}
\end{figure}

In the case of $d=10$ we can see similar behavior with $d=5$ case. As illustrated by the Fig. \ref{fig8}, the effect of correction terms are increasing of internal energy. As previous case, the higher order correction with positive coefficient (dash dot orange curve of the Fig. \ref{fig8}) includes a minimum for the internal energy. In presence of higher order corrections there is a critical point for small $M$ where correction terms have no effect (see crossed point of black dotted and blue dashed line of the Fig. \ref{fig8}). There is also a minimum $M$ where internal energy exists.\\

\begin{figure}[h!]
 \begin{center}$
 \begin{array}{cccc}
\includegraphics[width=80 mm]{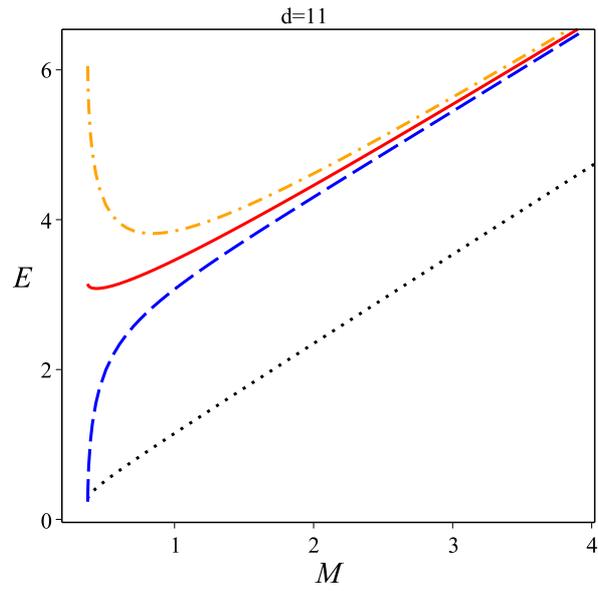}
 \end{array}$
 \end{center}
\caption{Internal energy in terms of $M$ for $d=11$.
$\alpha=0$ and $\gamma=0$ (black dot); $\alpha=1$ and $\gamma=0$ (red solid); $\alpha=1$ and $\gamma=-1$ (blue dash); $\alpha=1$ and $\gamma=1$ (orange dash dot).}
 \label{fig9}
\end{figure}

In the case of $d=11$, as previous case, we have large expression for the internal energy hence we give graphical analysis. We can see similar behavior with $d=5$ and $d=10$ cases. It is illustrated by the Fig. \ref{fig9}. It is clear that the effect of correction terms are increasing of internal energy. For the large $M$ higher order correction effect is negligible which is expected for massive (large) black hole.\\

\begin{figure}[h!]
 \begin{center}$
 \begin{array}{cccc}
\includegraphics[width=80 mm]{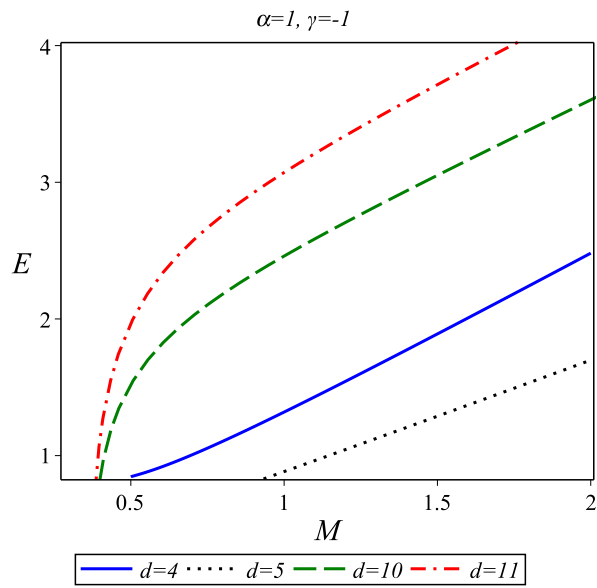}
 \end{array}$
 \end{center}
\caption{Internal energy in terms of $M$ for different dimensions.}
 \label{fig10}
\end{figure}

In order to compare corrected internal energy in several dimensions we give plot of the Fig. \ref{fig10}.

\subsection{Enthalpy}
\begin{figure}[h!]
 \begin{center}$
 \begin{array}{cccc}
\includegraphics[width=80 mm]{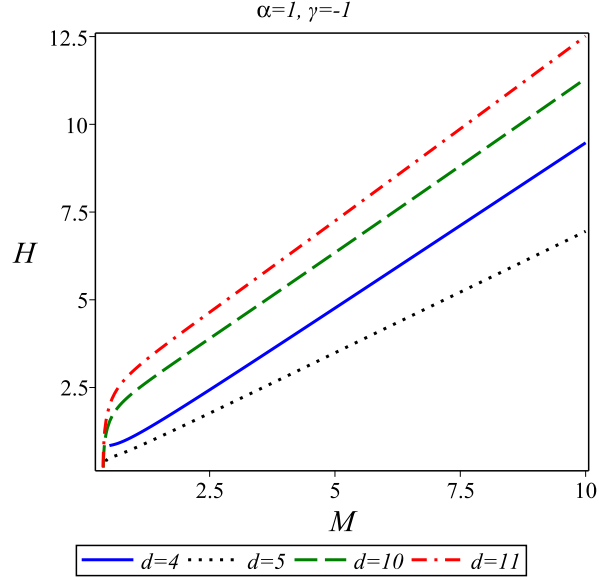}
 \end{array}$
 \end{center}
\caption{Enthalpy in terms of $M$ for different dimensions.}
 \label{fig11}
\end{figure}

In order to compute enthalpy we need black hole volume and pressure. Black hole volume is give by the following relation,
\begin{equation}\label{26}
V\propto r_{+}^{d-1}.
\end{equation}
Also, pressure obtained by using the following thermodynamics relation,
\begin{equation}\label{27}
P=-\frac{\partial F}{\partial V},
\end{equation}
where we should use event horizon $r_{+}$ given by the relation (\ref{18}). We find that effect of correction terms on the pressure at $d=4$ and $d=5$ are infinitesimal, while their effects at the $d=10$ and $d=11$ are non-trivial and may help to obtain critical points and dual Van der Waals behavior which will be discussed later.
Then, the enthalpy is given by using the following thermodynamics equation,
\begin{equation}\label{28}
H=E+PV.
\end{equation}
We find that effects of correction terms are infinitesimal at lower dimensions. In order to compare corrected enthalpy in several dimensions we give plot of the Fig. \ref{fig11}, and see that enthalpy is increasing function of $M$. Also, the enthalpy at $d=4$ and $d=5$ have the same behavior, while cases of $d=10$ and $d=11$ have similar manner. However, for the infinitesimal $M$ all cases yields to the same value. Also very large value of $M$ yields to vanishing effects of corrected terms as expected.\\
We have shown that the minimum value of enthalpy obtained at five dimensional space-time, which is illustrated by dotted black line of the Fig. \ref{fig11}.\\

\subsection{Gibbs free energy}
\begin{figure}[h!]
 \begin{center}$
 \begin{array}{cccc}
\includegraphics[width=70 mm]{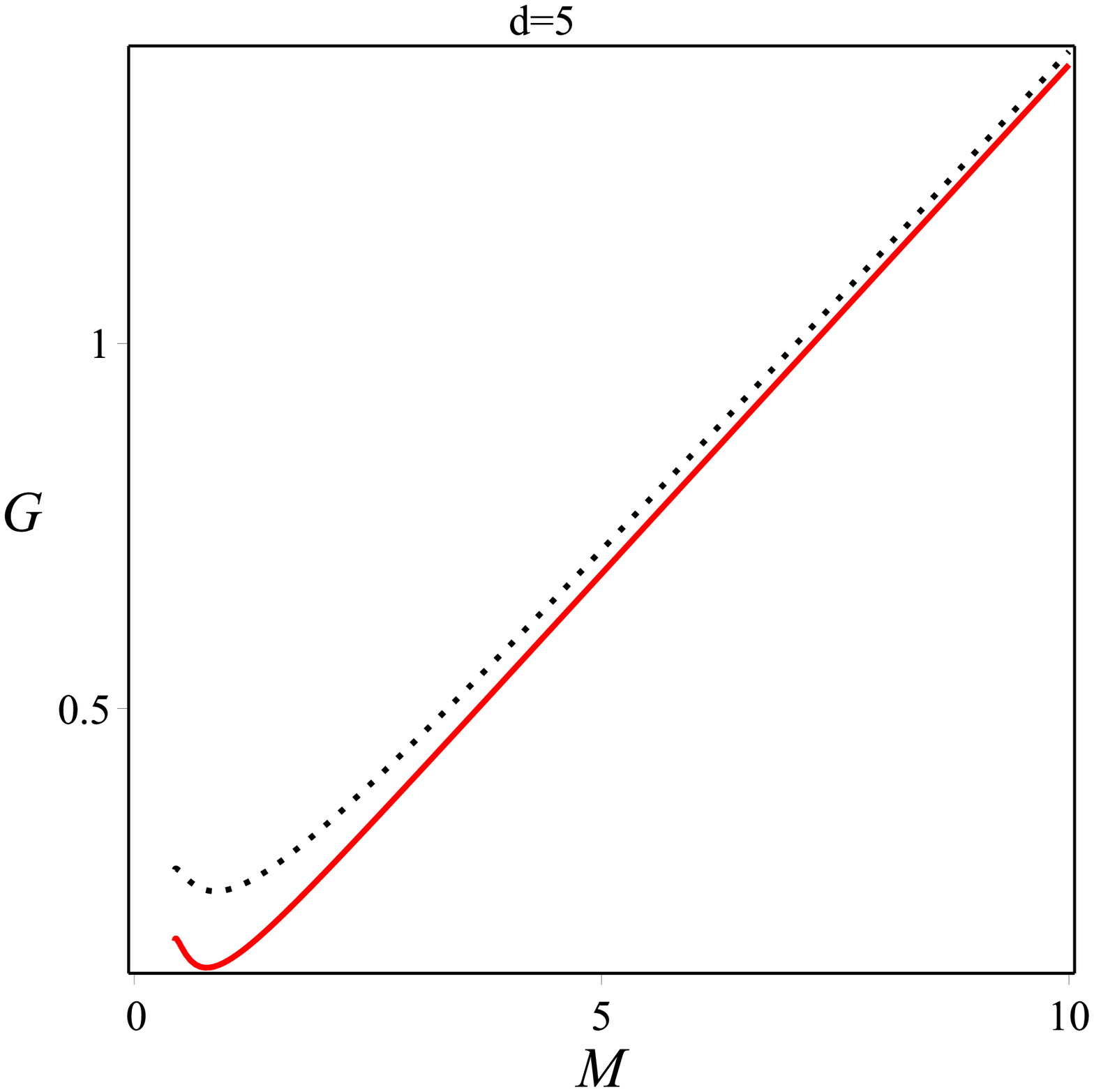}\includegraphics[width=70 mm]{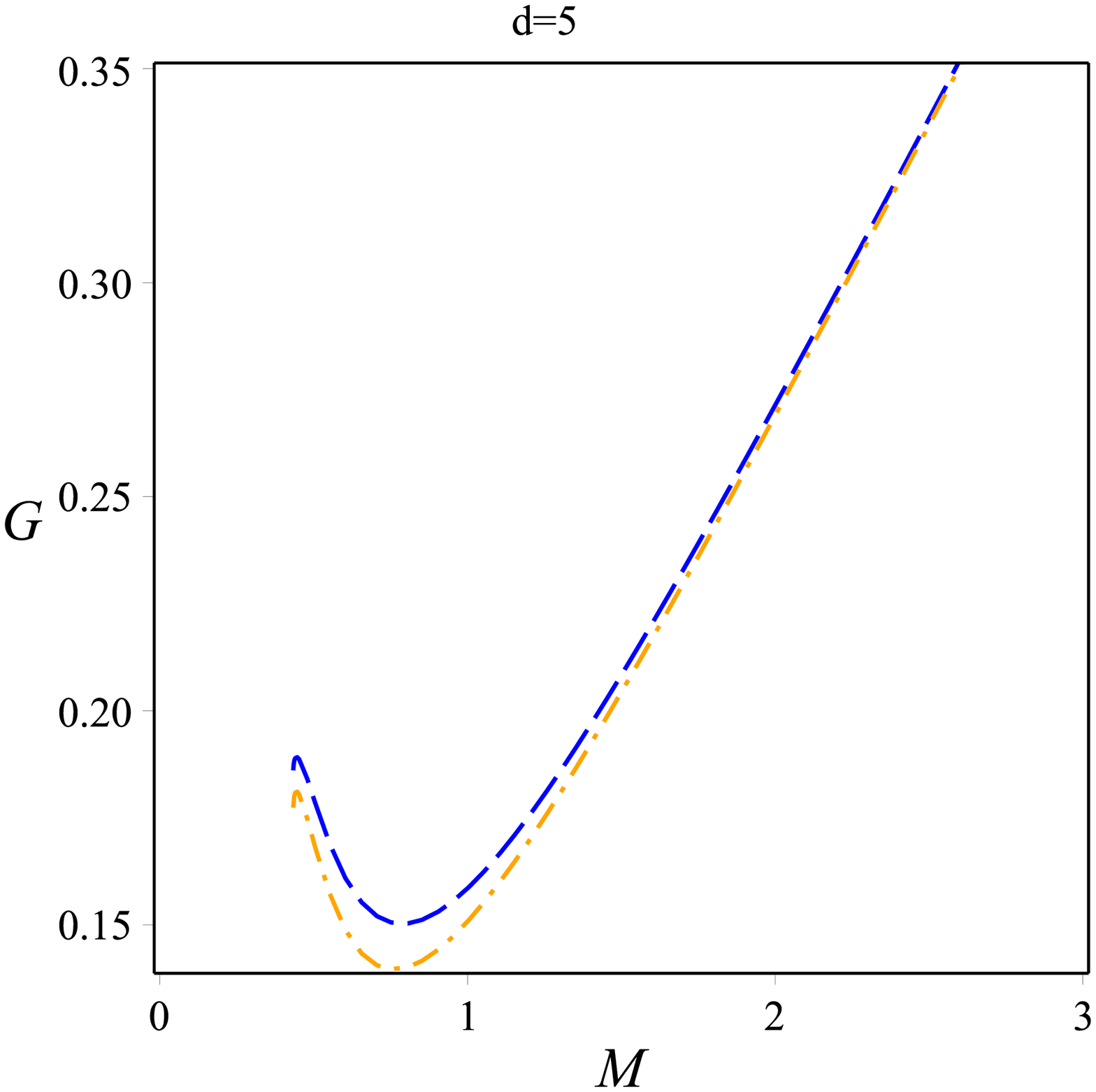}
 \end{array}$
 \end{center}
\caption{Gibbs free energy in terms of $M$ for $d=5$.
$\alpha=0$ and $\gamma=0$ (black dot); $\alpha=1$ and $\gamma=0$ (red solid); $\alpha=1$ and $\gamma=-1$ (blue dash); $\alpha=1$ and $\gamma=1$ (orange dash dot).}
 \label{fig12}
\end{figure}

By using enthalpy, which analyzed in the previous subsection, we can obtain another important thermodynamics quantity which is called Gibbs free energy and obtained using the following relation,
\begin{equation}\label{29}
G=H-TS.
\end{equation}
In the case of four dimensional space-time one can obtain the following expression,
\begin{eqnarray}\label{30}
G(d=4)&\approx&\frac{64M^{2}+3\pi^{2}}{12M\pi^{2}}\nonumber\\
&+&\frac{\alpha}{768\pi^{2}M^{3}}(256M^{2}-3\pi^{2})\ln{(128M^{2}-\pi^{2})}\nonumber\\
&-&\frac{\alpha}{384\pi^{2}M^{2}}(256M^{2}-3\pi^{2})\ln{M}\nonumber\\
&-&\frac{\alpha}{1536\pi^{2}M^{2}}\left((1024M^{2}-12\pi^{2})\ln{\pi}+(3840M^{2}-45\pi^{2})\ln{2}-2\pi^{2}\right)\nonumber\\
&-&\frac{\gamma}{30720\pi^{2}M^{5}}(160M^{2}-3\pi^{2}).
\end{eqnarray}
For the extra dimensions we give graphical analysis because Gibbs free energy expressions are so long and complicated also without correction.\\
So, in the plots of the Fig. \ref{fig12} we can see effects of logarithmic and higher order corrections on the Gibbs free energy. The left plots of the Fig. \ref{fig12} are include uncorrected (black dot) and logarithmic corrected (red solid) quantity in five dimensions. On the other hand right plot of the Fig. \ref{fig12} show positive and negative coefficient of higher order correction which its effects is infinitesimal.\\

\begin{figure}[h!]
 \begin{center}$
 \begin{array}{cccc}
\includegraphics[width=80 mm]{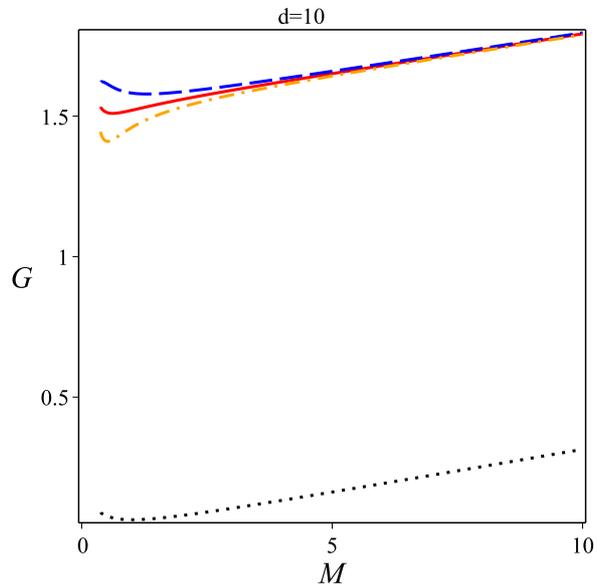}
 \end{array}$
 \end{center}
\caption{Gibbs free energy in terms of $M$ for $d=10$.
$\alpha=0$ and $\gamma=0$ (black dot); $\alpha=1$ and $\gamma=0$ (red solid); $\alpha=1$ and $\gamma=-1$ (blue dash); $\alpha=1$ and $\gamma=1$ (orange dash dot).}
 \label{fig13}
\end{figure}

Typical behavior of Gibbs free energy in ten dimensions plotted in the Fig. \ref{fig13}. We can see that thermal fluctuations increases value of the Gibbs free energy. Most difference between leading and higher order corrections is at small $M$. For the larger $M$ Gibbs free energy has linear behavior with respect to $M$.\\

\begin{figure}[h!]
 \begin{center}$
 \begin{array}{cccc}
\includegraphics[width=80 mm]{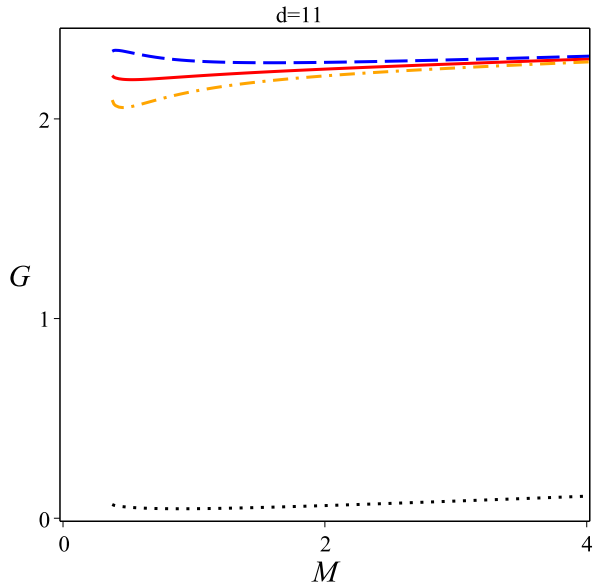}
 \end{array}$
 \end{center}
\caption{Gibbs free energy in terms of $M$ for $d=11$.
$\alpha=0$ and $\gamma=0$ (black dot); $\alpha=1$ and $\gamma=0$ (red solid); $\alpha=1$ and $\gamma=-1$ (blue dash); $\alpha=1$ and $\gamma=1$ (orange dash dot).}
 \label{fig14}
\end{figure}

General behavior of Gibbs free energy in eleven dimensions given by the Fig. \ref{fig14}. We can see that thermal fluctuations increases value of the Gibbs free energy which is similar to the case of ten dimensions. Most difference between leading and higher order corrections is at small $M$. For the larger $M$ Gibbs free energy yields to a constant which is different with the previous cases.\\

\begin{figure}[h!]
 \begin{center}$
 \begin{array}{cccc}
\includegraphics[width=80 mm]{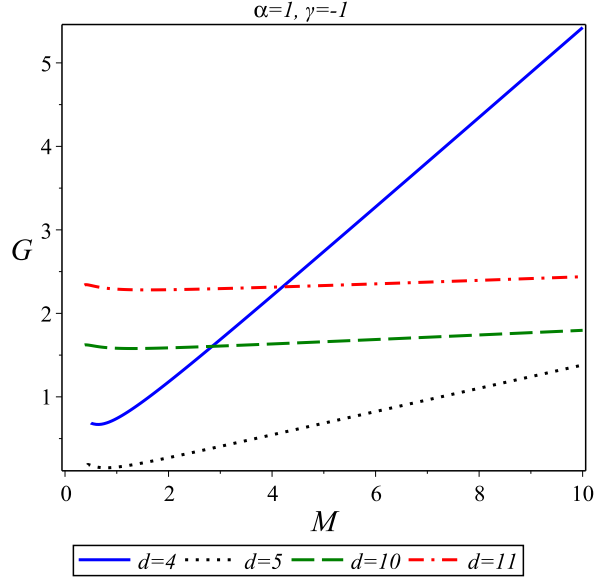}
 \end{array}$
 \end{center}
\caption{Gibbs free energy in terms of $M$ for different dimensions.}
 \label{fig15}
\end{figure}

In order to compare all cases of corrected Gibbs free energy we draw the Fig. \ref{fig15}. In the four-dimensional case, (blue  solid line of the Fig. \ref{fig15}) we can see that higher order corrected Gibbs free energy has linear behavior with the black hole mass.\\
black dotted line of the Fig. \ref{fig15} shows that the minimum value of the higher order corrected Gibbs free energy (for the intermediate mass) is belong to the five-dimensional case, which has linear behavior for the larger $M$.\\
In the cases of ten and eleven dimensions we can see that higher order corrected Gibbs free energy are approximately constant by variation of $M$. As the previous cases, we can see minimum value for the black hole mass to have Gibbs free energy.\\
There are critical points where value of the higher order corrected Gibbs free energy at four and ten dimensions as well as eleven dimensions are the same.\\
However, for the very larger $M$, higher order corrected Gibbs free energy of the five-dimensional case become bigger than the cases of ten and eleven dimensions.\\
We can see a minimum for the small mass limit of all case, hence it may be sign of some critical points and  stability for the small black hole mass.

\section{The first law of thermodynamics}
\begin{figure}[h!]
 \begin{center}$
 \begin{array}{cccc}
\includegraphics[width=80 mm]{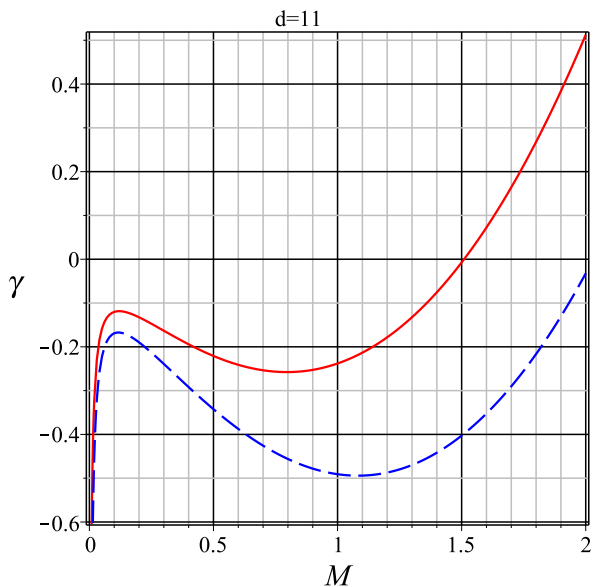}
 \end{array}$
 \end{center}
\caption{constraint of $\gamma$ in terms of $M$ for eleven dimensions. $\alpha=1$ denoted by red solid line while $\alpha=1.4$ denoted by blue dashed line.}
 \label{fig16}
\end{figure}

The first law of thermodynamics for the charged black hole given in this paper written as,
\begin{equation}\label{31}
dM=T dS+\Phi dQ,
\end{equation}
where
\begin{equation}\label{32}
\Phi=\frac{Q}{r_{+}^{d-3}},
\end{equation}
is the electrical potential and plays role of chemical potential in the dual field theory. Now, we would like to study above law for the higher dimensional Rissner-Nordstr\"{o}m black hole in presence of higher order corrections of the entropy.\\
It is clear that the first law is violated for the uncorrected thermodynamics which reflect some instabilities, however it is possible to remove such instabilities in presence of higher order corrections.\\
In the case of four dimensional case, one can obtain following condition to satisfy the first law of thermodynamics,
\begin{equation}\label{33}
\gamma\approx-\frac{32M^{2}\left([512\pi^{2}-4096]M^{4}+32\pi^{2}M^{2}+\alpha\pi^{2}\right)}{128M^{2}-\pi^{2}}.
\end{equation}
In the case of five dimensional case, one can obtain following condition to satisfy the first law of thermodynamics,
\begin{equation}\label{34}
\gamma\approx-\frac{3150M^{4}(M-1.5)}{136M^{2}-7}-\frac{\alpha}{2}\frac{353M^{4}+122M^{2}-15}{\sqrt{M}(136M^{2}-7)}.
\end{equation}
One can check that for some values of $M$ corrected parameter obtained as $\alpha=1$ and $\gamma=-1$.\\
In the case of ten dimensional case, one can obtain following condition to satisfy the first law of thermodynamics,
\begin{equation}\label{35}
\gamma\approx-\frac{3.2M^{\frac{8}{7}}}{(M^{2}-0.007)}(0.1M^{\frac{22}{7}}-0.007M^{\frac{8}{7}}-\alpha[0.2M^{2}-0.005]).
\end{equation}
In the case of eleven dimension we give graphical results as illustrated by the Fig. \ref{fig16}. We can see that for larger value of $M$, the first law of thermodynamics satisfied for the positive value of $\gamma$. Also there is some situations with $M=1$ (extremal case) where higher order correction vanished and the first law of thermodynamics satisfied by the logarithmic correction.

\section{Van der Waals equation}
\begin{figure}[h!]
 \begin{center}$
 \begin{array}{cccc}
\includegraphics[width=50 mm]{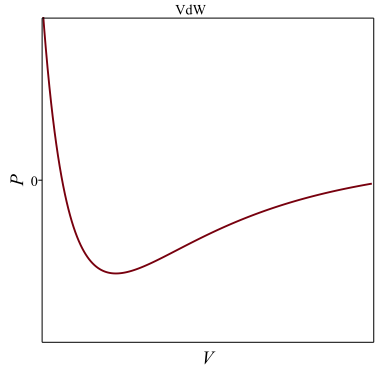}
 \end{array}$
 \end{center}
\caption{Van der Waals behavior.}
 \label{fig17}
\end{figure}

In this section we would like to investigate whether  the higher dimensional Rissner-Nordstr\"{o}m black hole in presence of higher order corrections of the entropy is holographically dual of a Van der Waals fluid or no. A Van der Waals fluid given by the following equation of state (in unit of Boltzmann constant),
\begin{equation}\label{36}
(P+\frac{a}{V^{2}})(V-b)=T,
\end{equation}
where $a$ and $b$ are some positive constants. In the Fig. \ref{fig17} we show typical behavior of the equation (\ref{36}) to compare with our results.\\

\begin{figure}[h!]
 \begin{center}$
 \begin{array}{cccc}
\includegraphics[width=65 mm]{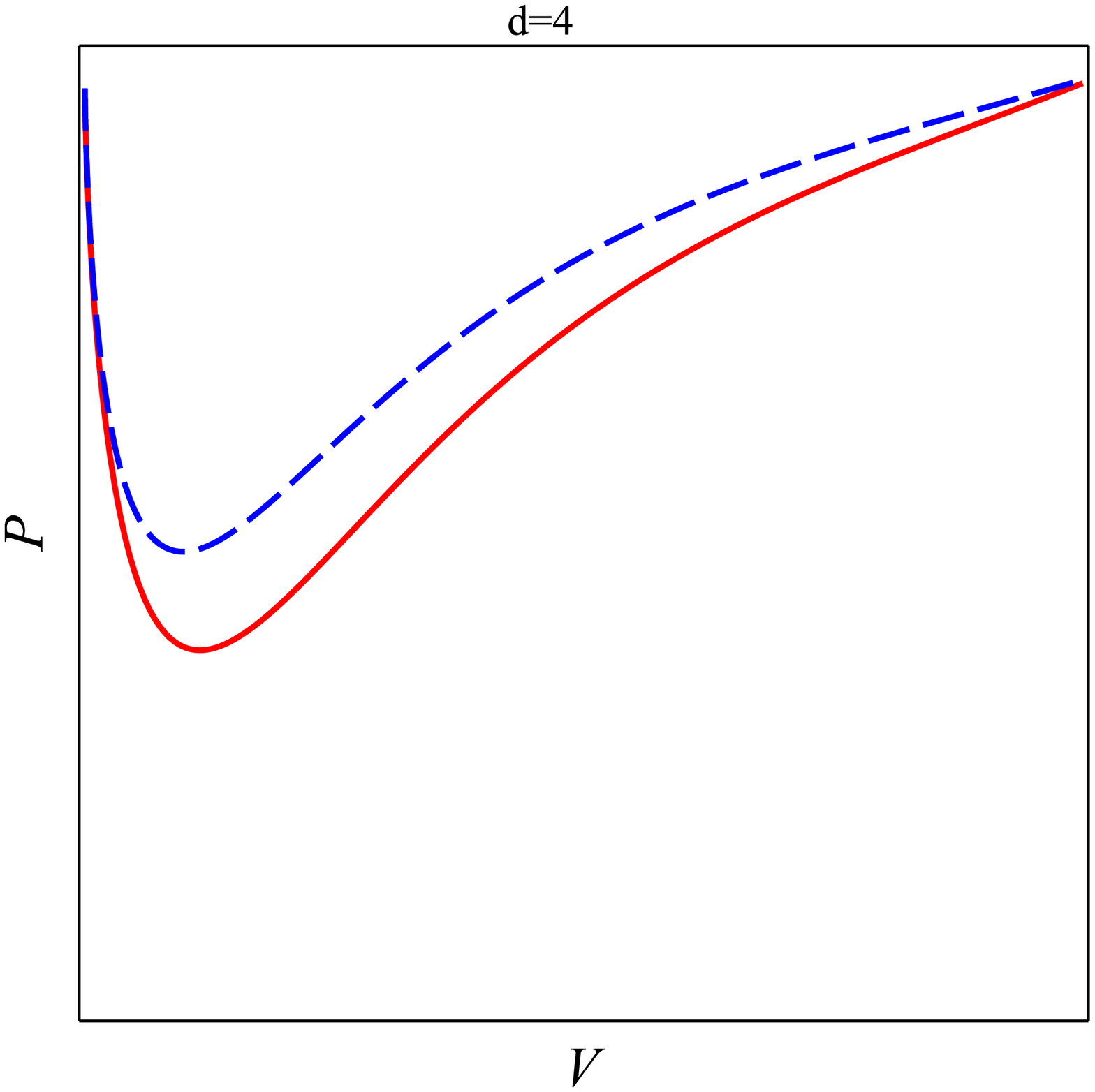}\includegraphics[width=65 mm]{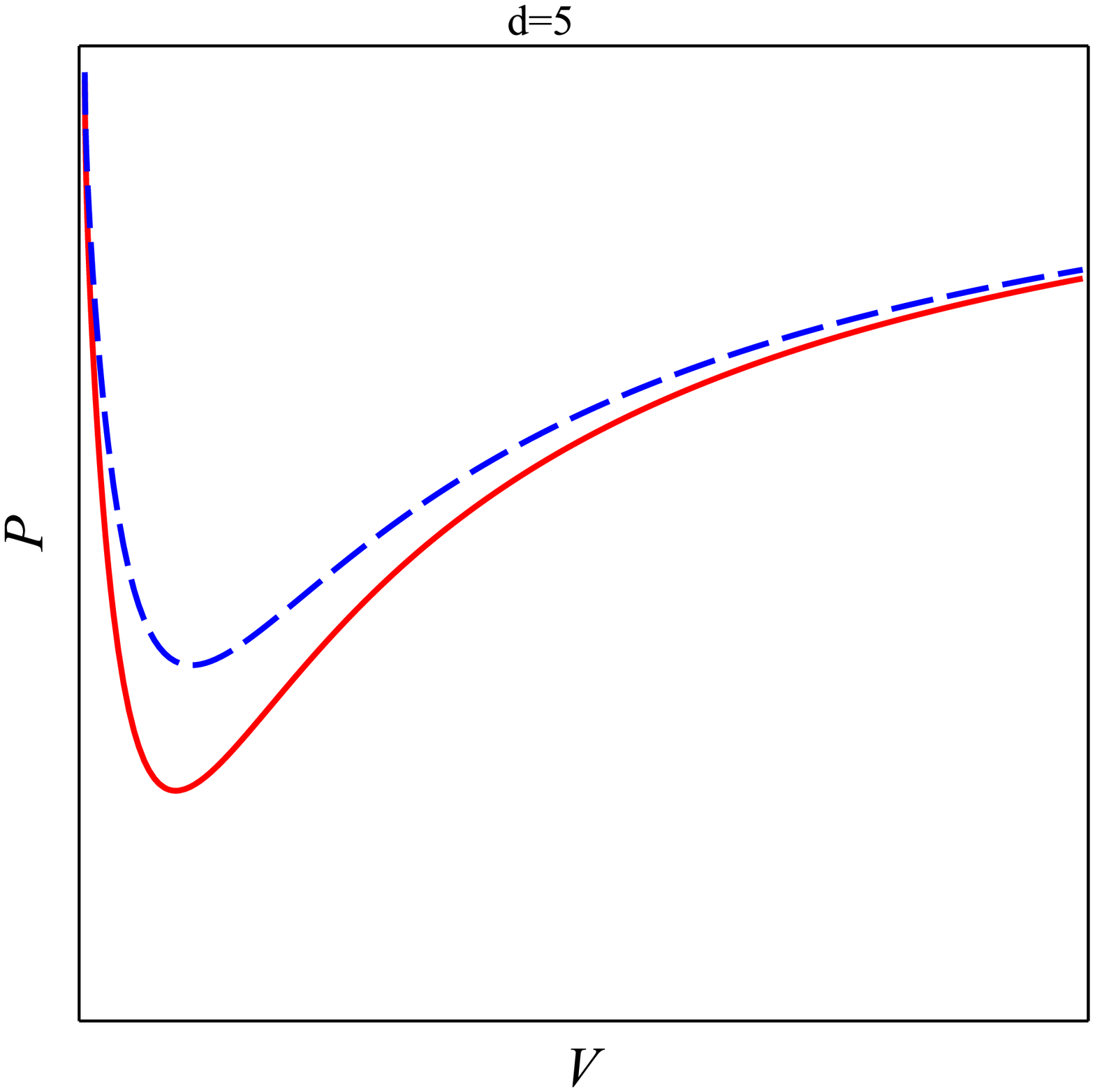}\\
\includegraphics[width=65 mm]{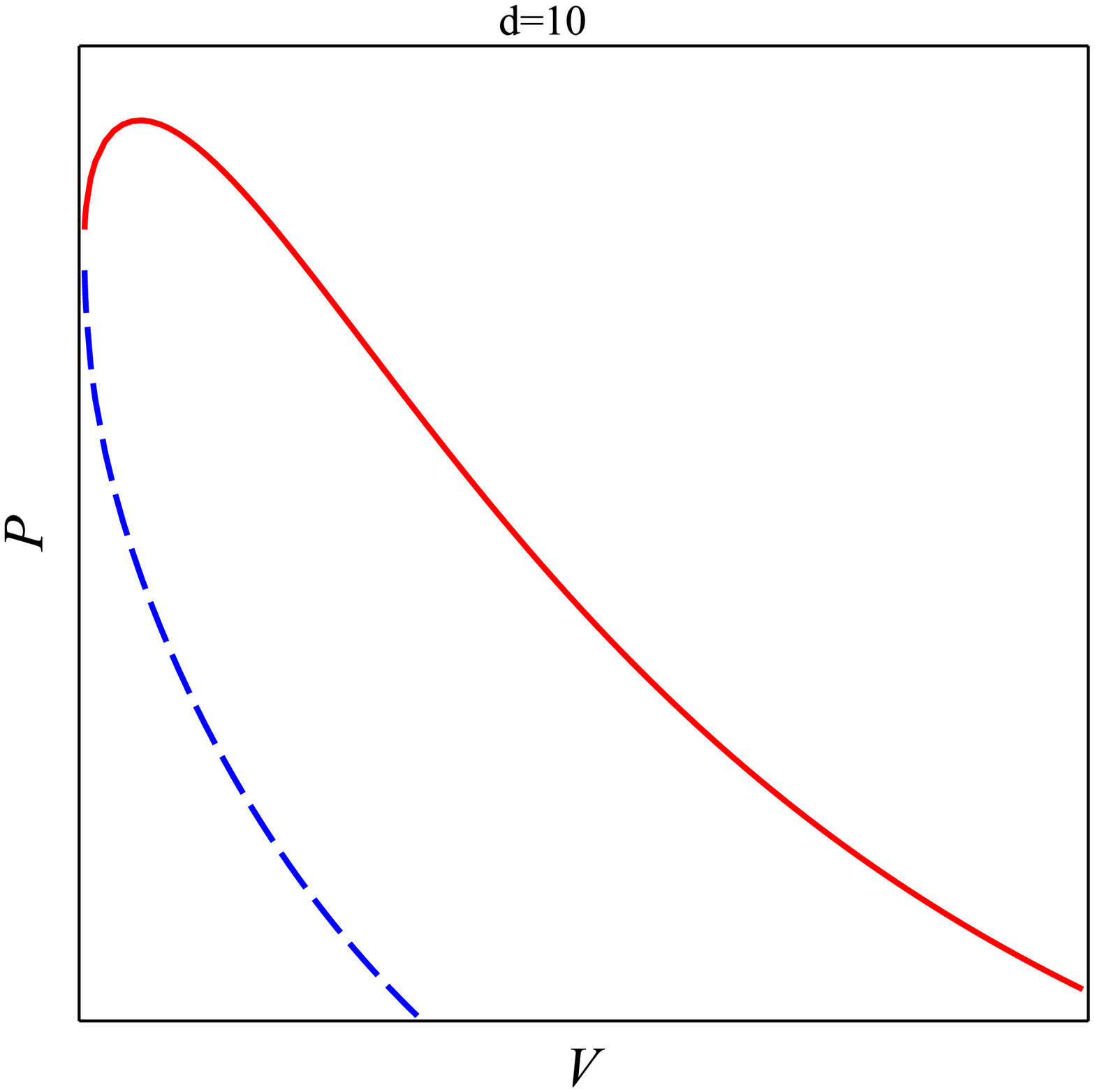}\includegraphics[width=65 mm]{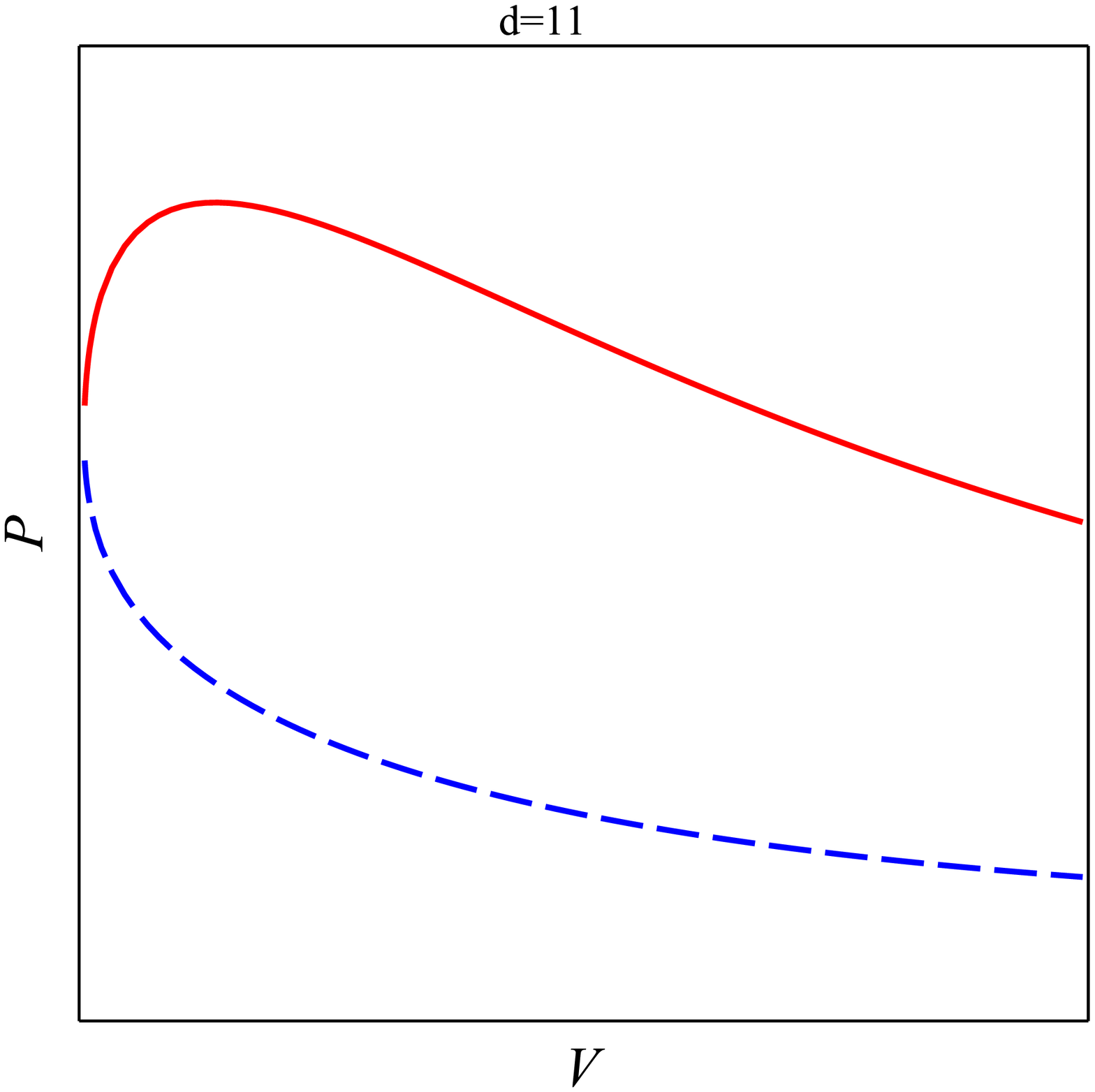}
 \end{array}$
 \end{center}
\caption{Pressure in terms of volume. Red solid line represents corrected case $\alpha=1$, $\gamma=-1$. Blue dashed line represents uncorrected case $\alpha=0$, $\gamma=0$.}
 \label{fig18}
\end{figure}

We find that for the both cases of corrected and uncorrected cases, there are Van der Waals like behavior for the lower dimensions while in the cases of $d=10$ and $d=11$ there are no such behavior. We show this by plots of the Fig. \ref{fig18} which are behavior of pressure in terms of volume. We find that effect of higher order correction is infinitesimal for the cases of $d=4$ and $d=5$. Upper plots of the Fig. \ref{fig18} are comparable with the Fig. \ref{fig17} which tells that Rissner-Nordstr\"{o}m black hole in four and five dimensions behave as Van der Waals fluid but lower plots have different behavior.

\section{Critical points and stability}
\begin{figure}[h!]
 \begin{center}$
 \begin{array}{cccc}
\includegraphics[width=60 mm]{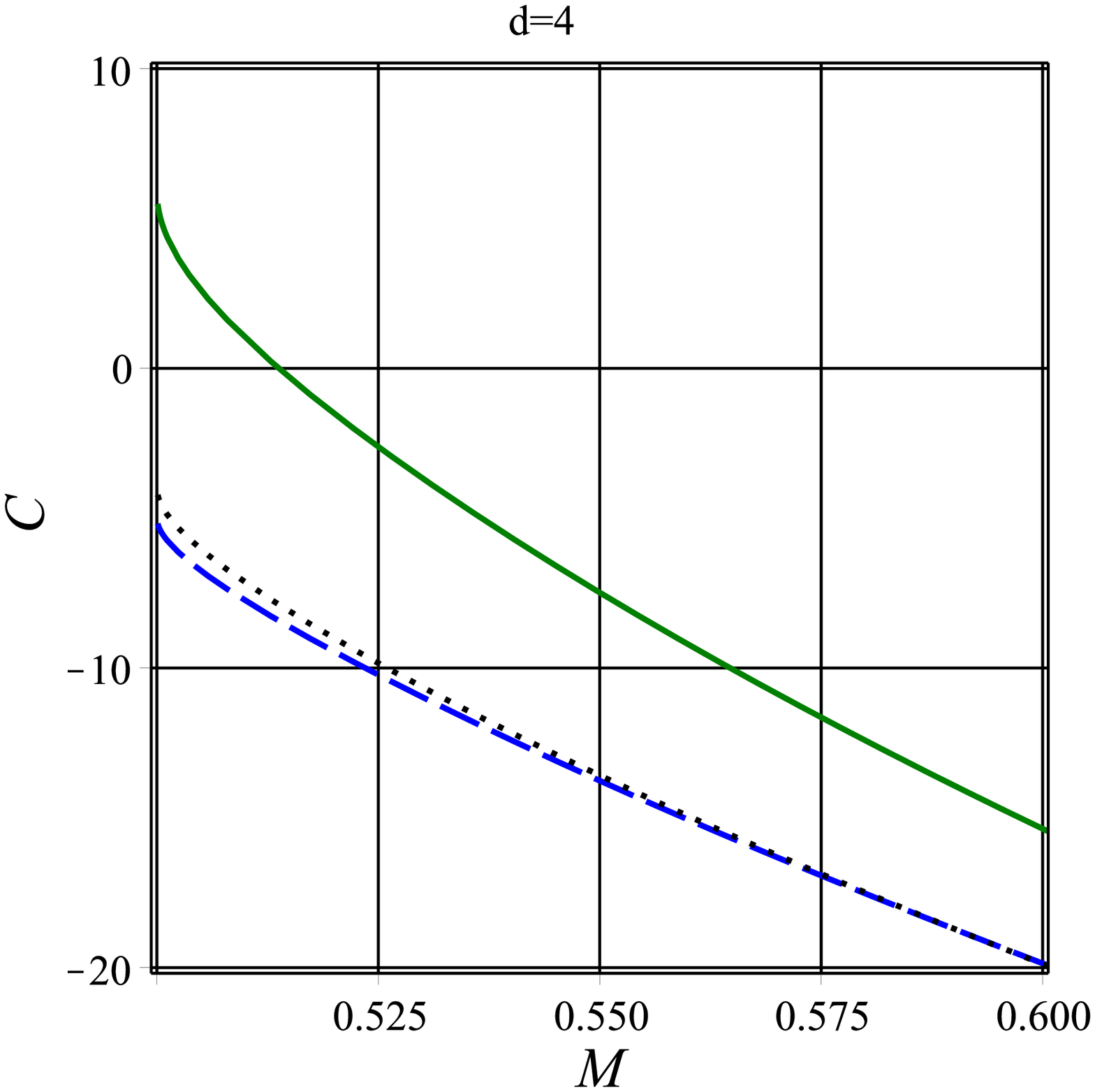}\includegraphics[width=60 mm]{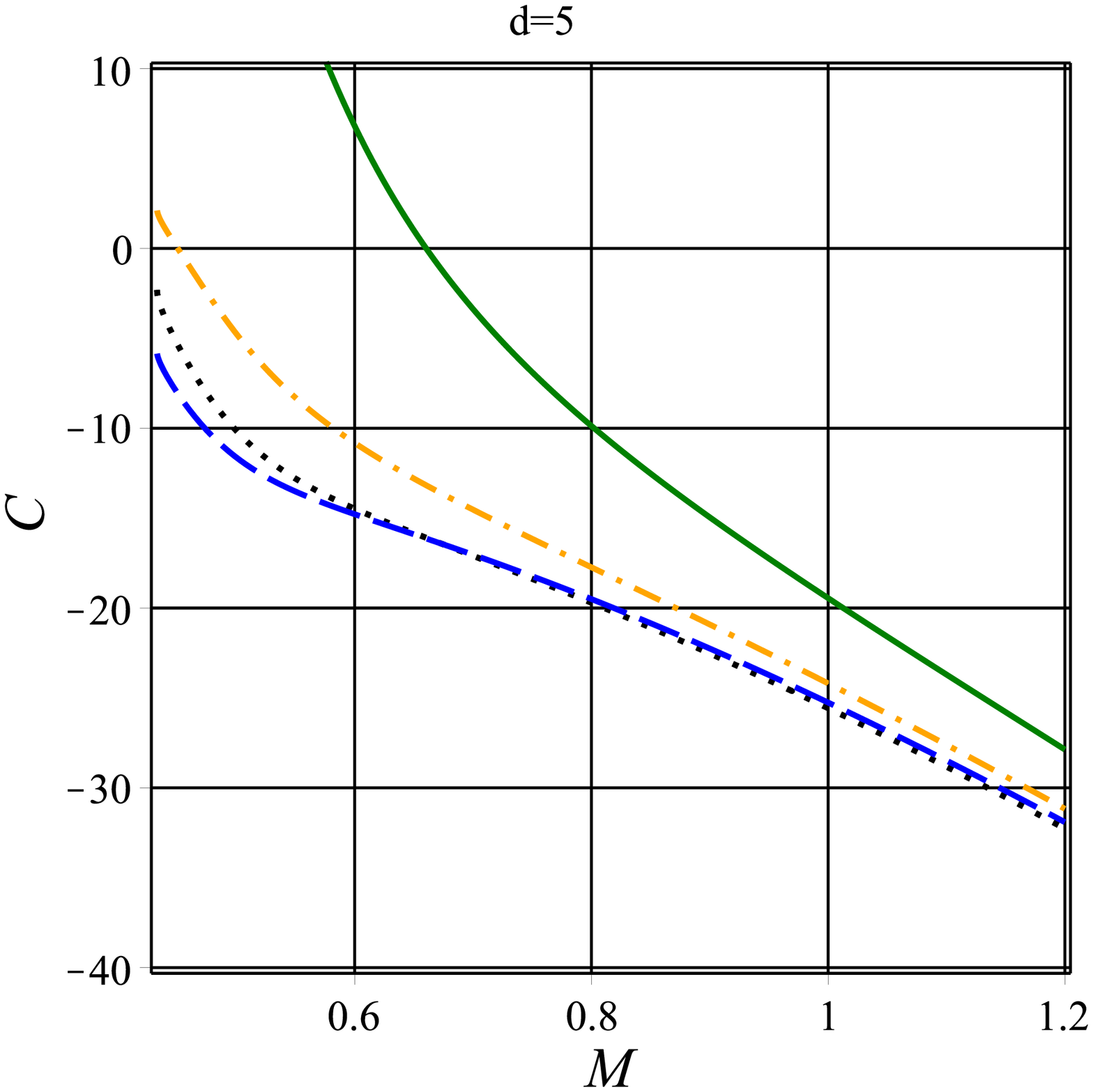}\\
\includegraphics[width=60 mm]{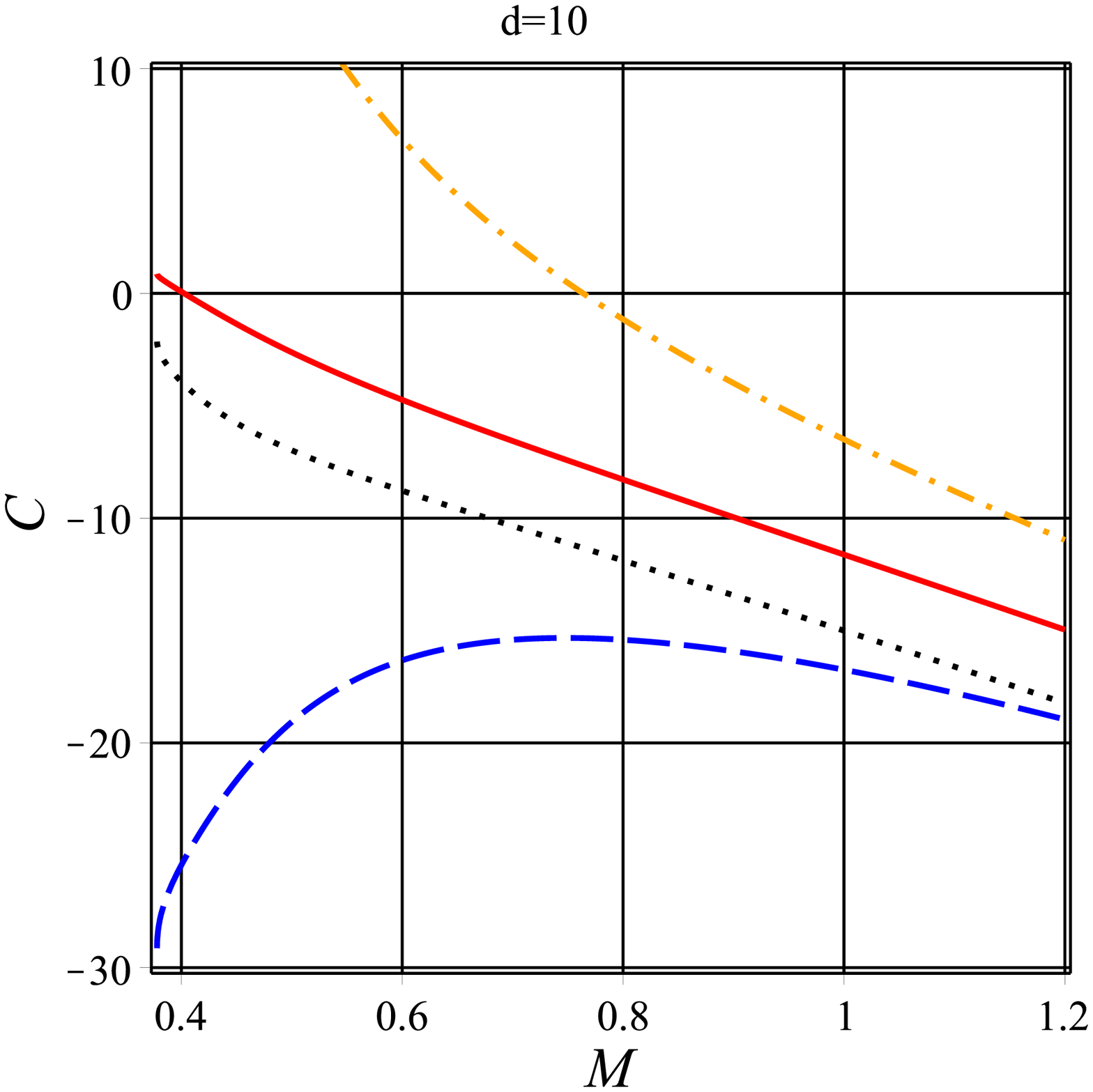}\includegraphics[width=60 mm]{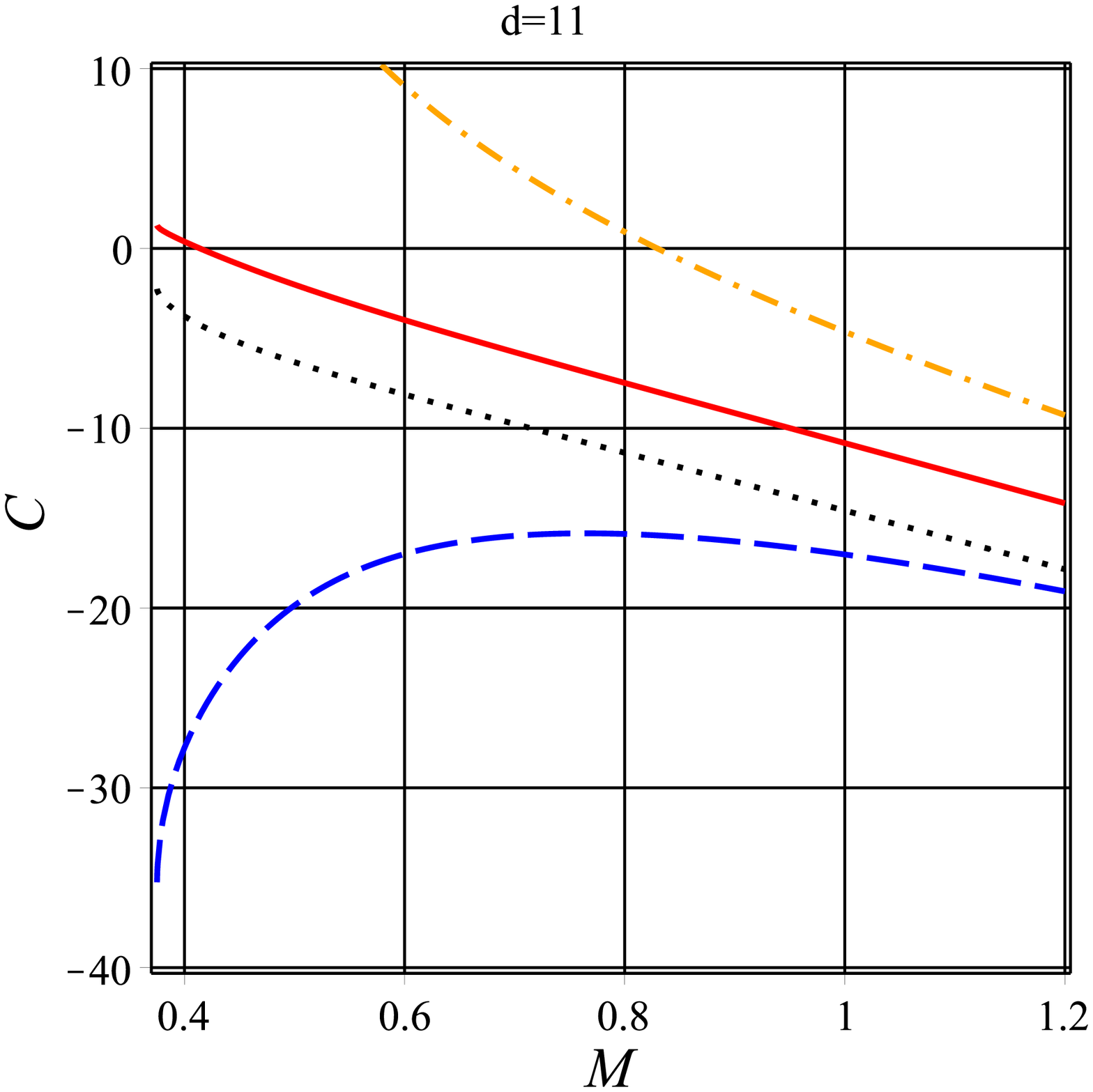}
 \end{array}$
 \end{center}
\caption{Specific heat in terms of $M$ for several space-time dimensions.
$\alpha=0$ and $\gamma=0$ (black dot); $\alpha=1$ and $\gamma=-1$ (blue dash); $\alpha=1$ and $\gamma=0$ (red solid); $\alpha=1$ and $\gamma=1$ (orange dash dot); $\alpha=1$ and $\gamma=1$ (green solid).}
 \label{fig19}
\end{figure}

Having pressure in terms of volume one can investigate critical points where satisfy following conditions simultaneously \cite{GRG3},
\begin{eqnarray}\label{37}
\frac{dP}{dV}=0,\nonumber\\
\frac{d^{2}P}{dV^{2}}=0.
\end{eqnarray}
We find that there is no critical point for the uncorrected case nor higher order corrected case.\\
We can also study stability of the model by using sign of specific heat given by the following relation,
\begin{equation}\label{38}
C=T(\frac{\partial S}{\partial T}).
\end{equation}
Graphical analysis of specific heat presented by the Fig. \ref{fig19}.
We find that presence of correction terms are necessary to have the black hole stability. In absence of thermal fluctuations, Rissner-Nordstr\"{o}m black holes are unstable. In the case of four dimensions, higher order corrected Rissner-Nordstr\"{o}m black hole will be stable at small $M$ for sufficiently large positive value of $\gamma$. It means that logarithmic correction has no any important effects on the black hole stability, and our interesting case of $\alpha=1$ and $\gamma=-1$ is also yields to unstable black hole. We should note that there is no phase transition and critical points in this case.\\
We obtain similar results for the five dimensional case. Higher order corrected Rissner-Nordstr\"{o}m black hole at five dimensions will be stable at small $M$ for positive value of $\gamma$, while negative $\gamma$ yields to unstable black hole without any phase transition and critical points.\\
Logarithmic corrected entropy will be affect stability of black hole at higher dimensions space-time. Lower plots of the Fig. \ref{fig19} show that ten and eleven dimensional cases yields to the stable black hole for small values of $M$. In all situations, our interesting case of $\alpha=1$ and $\gamma=-1$ yields to the unstable black hole. However, positive value of higher order corrected parameter yields to the stability of black hole. In all cases, large value of $M$ yields to the unstable black hole.

\section{Conclusion}
In this work, we considered higher dimensional Rissner-Nordstr\"{o}m black hole and investigated the effects of thermal fluctuations on the thermodynamics of the black holes at four, five, ten and eleven dimensions. Higher dimensional case selected inspired by AdS/CFT, superstring theory and M-theory respectively. The main point of this paper is consideration of higher order corrections of entropy which their origins presented in the section 2. We obtained effects of leading order (logarithmic correction) and higher order corrections of the entropy on the Helmholtz, internal, and Gibbs energies and also enthalpy. We found that four and five dimensional cases have similar behavior, while ten and eleven dimensions have similar behavior. We obtained suitable conditions on the higher order correction coefficient in several dimensions to satisfy the first law of thermodynamics. We have shown that without thermal fluctuations, black hole violates the first law of thermodynamics and yields to unstable black hole. Analyzing Van der Waals like behavior, show that there are no critical points and higher dimensional black hole $d>5$ does not behaves as Van der Waals fluid, while the cases of $d=4$ and $d=5$ may be holographically dual of Van der Waals fluid. By using such duality one can investigate quantum gravity effects coming from thermal fluctuations. The last sector of this paper study stability of black hole by using sign of the specific heat. We have shown that presence of logarithmic correction may yields to some stable points at higher dimensional case ($d=10$ and $d=11$). We can use higher order corrections for the other black holes like Horava-Lifshitz black hole \cite{HL1, HL2}, several kinds of hairy black holes \cite{hairy1, hairy2, hairy3, hairy4}. In the Ref. \cite{pour444} statistical mechanics of a new regular black hole  have been studied which can be extended by using higher order corrections of the entropy. Moreover, it is interesting to consider such effects for some two dimensional black holes \cite{2D1, 2D2}. Finally, it is interesting to repeat calculations of this paper in the presence of exponential nonlinear electrodynamics \cite{GRG4}.

\end{document}